%% file: npback2ndfor_eng.tex
\documentclass[a4paper,12pt]{article}

\usepackage[T2A]{fontenc}
\usepackage[cp1251]{inputenc}
\usepackage[english]{babel}
\usepackage{graphicx}

\usepackage[centertags]{amsmath}
\usepackage{amsfonts}
\usepackage{amssymb}
\usepackage{amsmath}
\usepackage{newlfont}

\usepackage{multirow}
\usepackage{caption}
\usepackage{longtable}
\usepackage{array}
\usepackage{lscape}

\renewcommand{\arraystretch}{1.2} 

\newlength{\defbaselineskip}
\newcommand{\setlinespacing}[1]%
           {\setlength{\baselineskip}{#1 \defbaselineskip}}

\makeatletter
\def\redeflsection{\def\l@section{\@dottedtocline{1}{0em}{10em}}}
\renewcommand{\appendix}{\par%
  \setcounter{section}{0}%
  \setcounter{subsection}{0}%
  \renewcommand{\appendixname}{Приложение}%
  \def\sectionname{\appendixname}%
  \addtocontents{toc}{\protect\redeflsection}%
  \gdef\thesection{\@Alph\c@section\@arabic\c@table}%
}

\begin{document}

\title{\bfseries\boldmath Charge-exchange quasi-elastic process $nd\to p(nn)$ under $0^\circ$ in the frame of elastic $np\to np$ scattering to $180^\circ$}

\maketitle


\begin{center}
 \author{\bf R.\,A.~Shindin, D.\,K.~Guriev, A.\,N.~Livanov, I.\,P.~Yudin}
 \vskip 3mm
 {\small{\it VBLHEP JINR, 141980 Dubna, Russia}\\
 {\it E-mails: shindin@jinr.ru, gurdm@yahoo.com, livanov@jinr.ru, yudin@jinr.ru}}
\end{center}

\vskip 3mm

 \hfill Pacs: {25.40.Kv}

 \hfill UDC: {539.171.11}

\vskip 3mm





\if 4
 \begin{abstract}
  Рассмотрена проблема спиновой физики,
  связанная с разницей представлений упругого взаимодействия
  между нейтроном и протоном.
  Взаимодействие можно назвать 
  процессом упругой $np\to pn$ перезарядки,
  либо расценить как реакцию $np\to np$ рассеяния нейтрона.
  Преобразование от одного представления к другому 
  обеспечивает унитарный оператор Майораны.
  В рамках импульсного преближения 
  дважды определена квазиупругая реакция перезарядки нейтрона на дейтроне.
  В представлении $nd\to p(nn)$ при рассеянии протона на угол $\theta$
  получена формула Дина.
  Используя представление $nd\to (nn)p$ квазиупругого рассеяния нейтрона
  (образующего с нейтроном-спектатором $nn$-пару)
  под углом $\pi-\theta$, определена альтернативная формула.
 \end{abstract}
\fi


 \begin{abstract}
  It is considered the problem of spin physics 
  related with the difference of representation of the elastic interaction 
  between the neutron and proton. 
  In the first case the charge-exchange reaction $np\to pn$ 
  under the angle $\theta$ is supposed, 
  in the second --- the simple elastic scattering of $np\to np$, 
  when the neutron is going in opposite direction $\pi-\theta$. 
  The transition from one representation to another
  is provided by the Majorana operator. 
  In the framework of impulse approximation 
  it is twice calculated 
  the quasi-elastic charge-exchange reaction of a neutron on a deuteron. 
  In the frame of $nd\to p(nn)$ scattering of proton to the angle $\theta$ 
  it gives the well-known Dean formula. 
  Using other representation $nd\to (nn)p$ 
  as a neutron elastic scattering under the angle $\pi-\theta$
  (together with neutron-spectator in $nn$-pair)
  the alternative formula is presented.
 \end{abstract}
 



\section{Introduction}
  The formalism of $NN$-interaction repeats
  the method intended to discribe the electron scattering
  on the particles with half-integer spin, for example,
  on other electron or on the atom with one electron
  outside of closed shell. Therefore in the beginning
  the definitions of spin and space fermion funtions
  will be considered.

   The spin-vector is an analog of mechanical moment
   but its projection to any direction has a discrete values.
   In the case of particle with spin $\frac12\hbar$
   the projection will be equal to $s_z=+\frac12\hbar$ or $-\frac12\hbar$.
   To account this duality 
   the three Hermitian $2\times2$ matrices are used 
   that forms the Pauli-operator $\widehat\sigma$:
  \begin{eqnarray*}\label{theory.Spin Pauli operator}
   & \widehat\sigma = \vec{i}\sigma_x + \vec{i}\sigma_y
   + \vec{k}\sigma_z \;,\quad\textrm{where} \\ 
   & \sigma_x = \begin{pmatrix}0&1\\1&0\end{pmatrix}\;,\quad
     \sigma_y = \begin{pmatrix}0&-i\\i&0\end{pmatrix}\;,\quad
     \sigma_z = \begin{pmatrix}1&0\\0&-1\end{pmatrix}\;.
  \end{eqnarray*}

  Since only $\sigma_z$ has a diagonal view 
  its own vectors or spinors are pure states:  
  \begin{eqnarray*}\label{theory.pure spinors}
   & \chi_z(s_z=+\tfrac12\hbar)=\dbinom10\;,\quad
     \chi_z(s_z=-\tfrac12\hbar)=\dbinom01\;.
  \end{eqnarray*}

\begin{figure}[!ht]
   \quad\centering
   \scalebox{.3}{\includegraphics{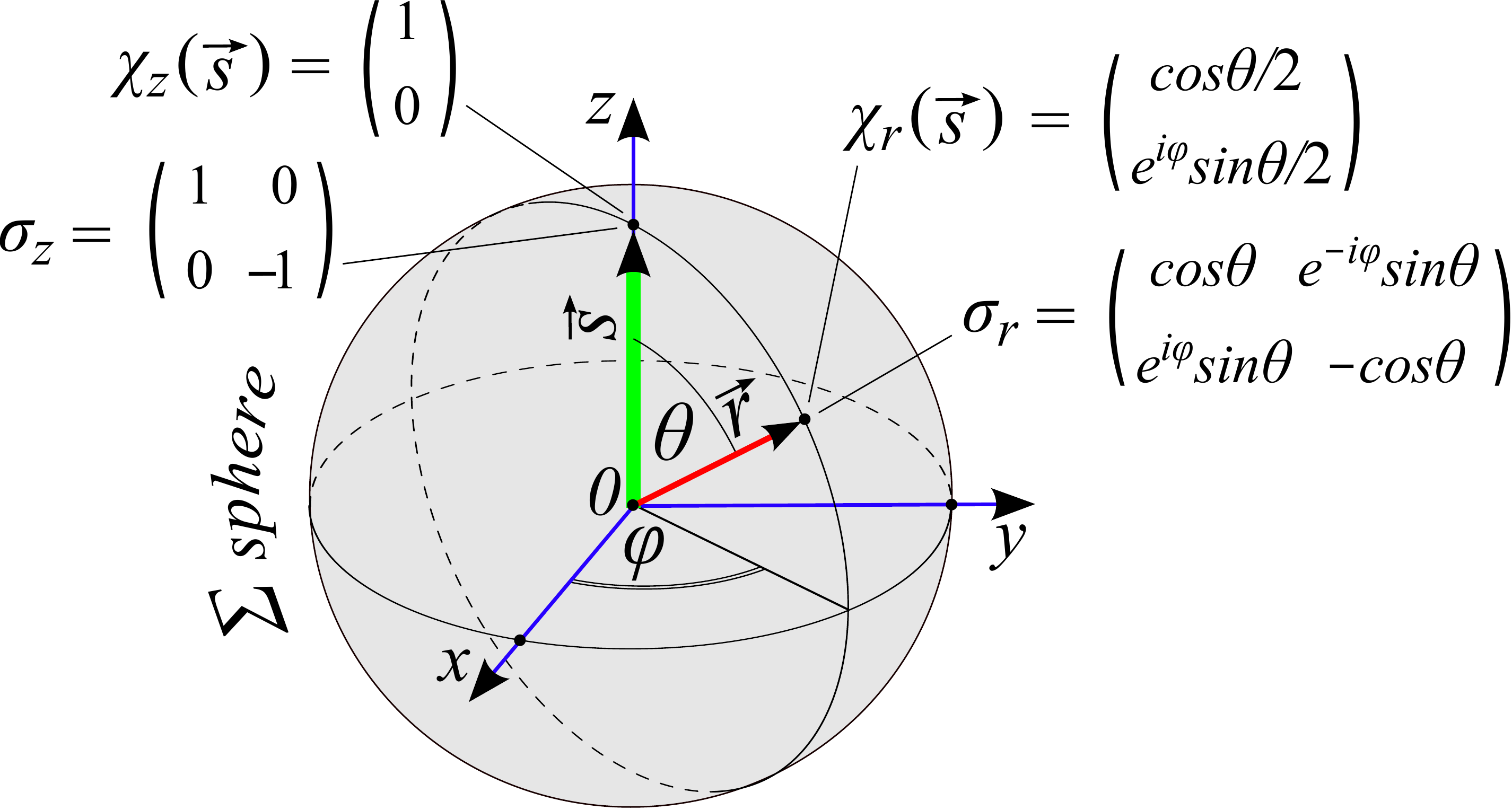}}\quad
   \caption{\small{Projection of the Pauli operator 
    $\widehat\sigma=\vec{i}\sigma_x+\vec{j}\sigma_y+\vec{k}\sigma_z$
    to any diection $\vec{r}$ equals to unit: $|\sigma_r|^2=1$.
    That allows to present its like a $\Sigma$-sphere.
    Each point of the $\Sigma$-sphere corresponds to operator $\sigma_r$
    which has two own spinors $\chi_r(s_z=+\tfrac{\hbar}{2})$
    and $\chi_r(s_z=-\tfrac{\hbar}{2})$.
   }}\label{theory.Pauli operator sphere}
  \end{figure}

\begin{figure}[!ht]
   \quad\centering
   \scalebox{.3}{\includegraphics{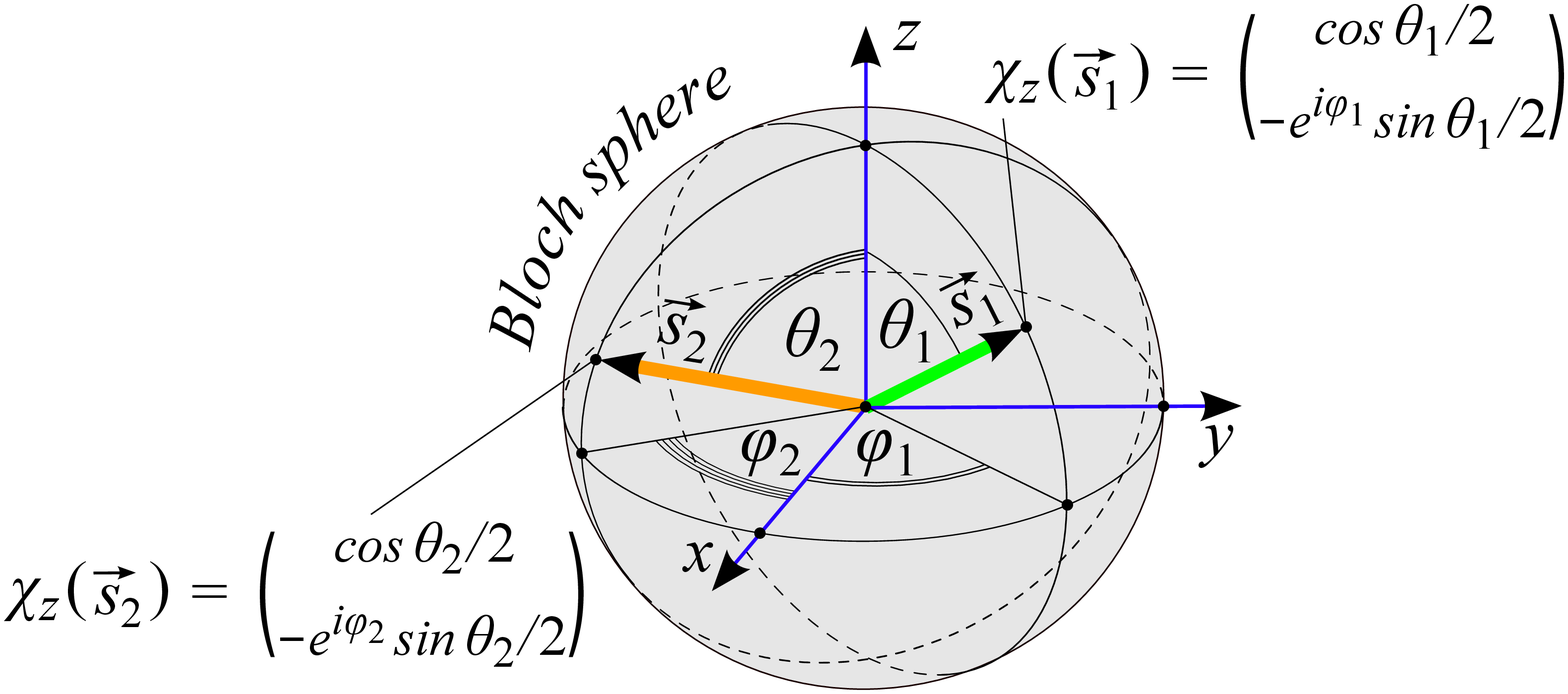}}\qquad
   \caption{\small{The Bloch sphere.
    The spin direction $\vec{s}\equiv\vec{s}(\theta,\varphi)$
    is defined by the angles $\theta$ and $\varphi$.
    For each $\vec{s}$ the point on the Bloch sphere
    corresponds to the single spin state along $z$-axis:
    $\chi_z(\vec{s})=\cos\frac\theta2\binom10-e^{i\varphi}\sin\frac\theta2\binom01$.
    The sign minus before low spinor element 
    arises due to the inverse count of angle $\theta$
    in comparison with the case on Fig.~\ref{theory.Pauli operator sphere}.
   }}\label{theory.Bloch sphere}
  \end{figure}

  For each fermion there is own polarization axis $\vec{s}$
  where its state is defined by the spinor $\binom10$.
  Along any other directions the fermion spin state will be mixed.
  Own vectors of matrix $\sigma_r=(\widehat\sigma\cdot\vec{r}\,)$
  (Fig.~\ref{theory.Pauli operator sphere})
  will have both components $\binom10$ and $\binom01$
  which define two variants of spin polarization.
  If fermion is polarized along $z$
  its sate along $\vec{r}$ expresses as follows:
  \begin{equation*}
    \chi_r(s_z=+\tfrac{\hbar}{2})=\binom{\cos\frac\theta2}{e^{i\varphi}\sin\frac\theta2}=
    \cos\frac\theta2\binom10+e^{i\varphi}\sin\frac\theta2\binom01\;.
  \end{equation*}
  Here $\cos\frac\theta2$ and $e^{i\varphi}\sin^2\frac\theta2$
  are the amplitudes of probability 
  to have the spin projections $s_r=+\tfrac12\hbar$
  and $-\tfrac12\hbar$.
  In our case there are two fermions and each of them 
  can have own spin direction.
  Therefore more sutable to use the definition of Bloch sphere
  where the spin states of particles are considered in one coordinate system 
  and quantized along $z$-axis (Fig.~\ref{theory.Bloch sphere}).
  
  \noindent
  \noindent

  To distinguish both particles the follows definition will be used:
  \begin{eqnarray}\label{theory.preliminary spin wave function of each fermion}
    \chi_{z,\,n}(\vec{s}_m)\;\equiv\;|\vec{s}_m\rangle_n\;=\;
    \binom{\cos\frac{\theta_m}2}{-e^{i\varphi_m}\sin\frac{\theta_m}2}_{\!\!n}
    \;,\quad n,\,m=\{1,2\}\;.
  \end{eqnarray}
  The index $m=1,\,2$ expresses the number of states.
  The order numbers of particles $n=1,\,2$ are arbitrary.
  Thus the vectors $|\vec{s}_1\rangle_1$ and $|\vec{s}_1\rangle_2$
  present one fermion
  but named its either first or second particle.
  
  In most cases so detailed definitions of spin states
  \eqref{theory.preliminary spin wave function of each fermion}
  are superfluous and enough to define them as spinors
  $\binom\alpha\beta$ and $\binom\gamma\delta$
  where the $\alpha,\,\beta,\,\gamma,\,\delta$ have complex values
  and satisfy to the conditions: $|\alpha|^2+|\beta|^2=|\gamma|^2+|\delta|^2=1$.
  Independence of particles leads to factorization of their function: $\chi_{12}=\chi_1\chi_2$.
  The freedom of order numbers provides the following:
  \begin{equation}\label{theory.spin wave function of two fermions}
    |\vec{s}_1,\vec{s}_2\rangle=\binom\alpha\beta_{\!\!1}\binom\gamma\delta_{\!\!2}\;,\quad
    |\vec{s}_2,\vec{s}_1\rangle=\binom\gamma\delta_{\!\!1}\binom\alpha\beta_{\!\!2}\;.
  \end{equation}

  To describe a wave properties of particle it is sutable to use the exponent
  $e^{\,i(\vec{p}\vec{r}-E t)/\hbar}$
  where $\vec{p}$ and $E$ --- momentum and full energy of particle,
  $\vec{r}$ and $t$ --- space and time variables.
  Such harmonic is an approximation and presents the infinite plane wave.
  For our puposes the stationary functions ($t=$ const) will be enough.
  To distinguish both particle the analogical
  \eqref{theory.preliminary spin wave function of each fermion}
  definitions are choosed:
  \begin{equation}\label{theory.preliminary space wave function of each particle}
    \varphi_{\vec{p}_m}(\vec{r}_n)\;\equiv\;|\vec{p}_m\rangle_n\;=\;
    Ce^{\tfrac{i}{\hbar}\,\vec{p}_m\vec{r}_n}\;,\quad n,\,m=\{1,2\}\;,\quad
    C = (2\pi\hbar)^{-\frac{3}{2}}\;.
  \end{equation}
  Momentum has own index $m$
  which does not depend from the number $n$.
  Thus the vectors $|\vec{p}_1\rangle_1$ and $|\vec{p}_2\rangle_1$
  show that first particle can have momentum $\vec{p}_1$
  either momentum $\vec{p}_2$.
  Functions
  \eqref{theory.preliminary space wave function of each particle}
  are orthogonally among themselves
  that expresses the independence of particles and provides the factorization:
  \begin{equation*}\label{theory.space wave function of two particles}
    |\vec{p}_1,\vec{p}_2\rangle=|\vec{p}_1\rangle_1\,|\vec{p}_2\rangle_2\;,\quad
    |\vec{p}_2,\vec{p}_1\rangle=|\vec{p}_2\rangle_1\,|\vec{p}_1\rangle_2\;.
  \end{equation*}

  In the c.m.s. the momenta permutation
  is equivalent to the conjegation:
  \begin{eqnarray}\label{theory.space wave function of two particles complex double}
    & |\vec{p}_1,\vec{p}_2\rangle\,\to\,\varphi_{12}=
      Ce^{\tfrac{i}{\hbar}\,\vec{p}\,\vec{r}}\;,\quad
      |\vec{p}_2,\vec{p}_1\rangle\,\to\,\varphi^*_{12}=
      Ce^{-\tfrac{i}{\hbar}\,\vec{p}\,\vec{r}}\;, \\
    & \textrm{where}\;\vec{r}=\vec{r}_1-\vec{r}_2\;\textrm{ --- relative radius-vector}. \nonumber
  \end{eqnarray}

\begin{figure}[!ht]
   \centering
   \scalebox{.33}{\includegraphics{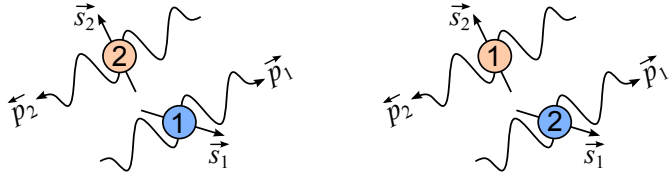}}
   \caption{\small{On the left the sates of two particles 1 and 2 
    are defined by the function
    $|\vec{p}_1,\vec{s}_1\rangle_1\,|\vec{p}_2,\vec{s}_2\rangle_2$.
    On the right the order numbers of partices are changed
    to correspond to the function
    $|\vec{p}_2,\vec{s}_2\rangle_1\,|\vec{p}_1,\vec{s}_1\rangle_2$.
   }}\label{theory.transition to antisimmetry}
  \end{figure}

  Acording to the Pauli principal the wave function $\Psi$
  of identical fermions should be antisymmetric
  relative the permutation of their spins and momenta:
  \begin{eqnarray}\label{theory.prime function modify}
    & \Psi=\dfrac{1}{\sqrt2}
    \Bigl(|\vec{p}_1,\vec{s}_1\rangle_1\,|\vec{p}_2,\vec{s}_2\rangle_2-
    |\vec{p}_2,\vec{s}_2\rangle_1\,|\vec{p}_1,\vec{s}_1\rangle_2\Bigr)\;, \\ 
    & \textrm{where}\;|\vec{p}_m,\vec{s}_m\rangle_n\,=\,|\vec{p}_m\rangle_n\,|\vec{s}_m\rangle_n
     \;. \nonumber
  \end{eqnarray}
  The full permutation don't change the particles configuration
  (Fig.~\ref{theory.transition to antisimmetry}).
  Thus the momentum of fermion with spin polarization $\vec{s}_1$
  anyway equals to $\vec{p}_1$.

\pagebreak
  The formulas \eqref{theory.spin wave function of two fermions}
  and \eqref{theory.space wave function of two particles complex double}
  allow to express the wave function \eqref{theory.prime function modify} as follows:
  \begin{equation}\label{theory.prime function modify 2}
    \Psi\;=\;\dfrac{1}{\sqrt2}
    \left[\dbinom\alpha\beta_1\dbinom\gamma\delta_2\varphi_{12}\;-\;
    \dbinom\gamma\delta_1\dbinom\alpha\beta_2\varphi^*_{12}\right]
    \textrm{\quad or\quad}
    \dbinom\alpha\beta_{p}\dbinom\gamma\delta_{-p}\;,
  \end{equation}
  \begin{equation}\label{theory.prime function modify 2 normalization}
    \int\int\Psi^+(\vec{p}\,')\Psi(\vec{p})\,d^3\vec{r}\,d^3\vec{p}\;=\;
    \int\delta(\vec{p}-\vec{p}\,')\,d^3\vec{p}\;=\;1\;.
  \end{equation}

  Taking into account the rule \eqref{theory.prime function modify 2 normalization}
  we will suppose for simplifiacation
  that the function \eqref{theory.prime function modify 2}
  is normalized to unit: $|\Psi|^2=1$.

\section{Fermions elastic scattering}\label{chapter-Theory.plane wave agriment}
\begin{figure}[!ht]
   \quad\centering
   \scalebox{.4}{\includegraphics{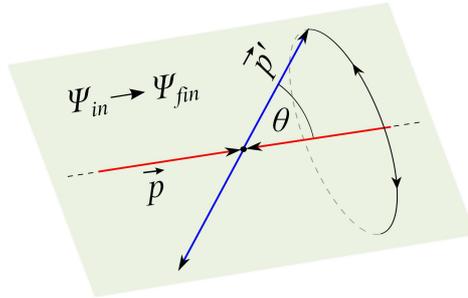}}\quad
   \caption{\small{The plane of interaction of two particles.
    Circular arrows show the freedom of rotation of this plane
    around the direction $\vec{p}$.
   }}\label{theory.plane wave}
  \end{figure}

  The result of interaction of two particles can be presented
  as a transformation of initial wave function $\Psi_{in}$ to final $\Psi_{fin}$
  (Fig.~\ref{theory.plane wave})
  and both of them are supposed as the plane waves
  \eqref{theory.prime function modify 2}.
  To change the momentum direction $\vec{p}\to\vec{p}\,'$
  let us to define the special unitary operator\footnote{The operator  
  $\widehat{P}(\theta)$ can be presented
  as $3\times3$ matrix $A$ which rotates 
  the momentum: $\vec{p}\, '= A\vec{p}$.
  The transformation $\widehat{P}(\theta)\varphi\to \varphi'$
  \eqref{theory.properties of turn operator}
  carry out in the c.m.s.
  If we go into the laboratory coordinates 
  the action of this operator reduces to multiplying 
  the wave function $|\vec{p}_1,\vec{p}_2\rangle$
  by the exponent $e^{-\tfrac{i}{\hbar}\vec{q}\,(\vec{r}_1-\vec{r}_2)}$
  where the $\vec{q}$ is a momentum of recoil particle.} $\widehat{P}(\theta)$:
  \vspace {-2mm}
  \begin{equation}\label{theory.properties of turn operator}
    \begin{array}{c}
      \widehat{P}(\theta)\times\varphi_{12} = \varphi\,_{12}' =
        e^{\tfrac{i}{\hbar}\vec{p}\,'\vec{r}}\;,\quad
      \cos\theta=\dfrac{(\vec{p},\vec{p}\,')}{|\vec{p}|\cdot|\vec{p}\,'|}\;, \\ [3mm]
      |\widehat{P}(\theta)|^2=1\;,\quad
      \widehat{P}(\theta)\widehat{P}(\omega)=\widehat{P}(\theta+\omega)\;,\quad
      \widehat{P}(\theta)=\widehat{P}(\theta+2\pi n)\;. \\ [3mm] 
      \Psi_{fin}\;=\;\widehat{P}(\theta)\times\Psi_{in}\;.
    \end{array}
  \end{equation}

  The momentum $\vec{p}\,'$ lies between the angles $\theta$ 
  and $\theta+d\theta$ on the reaction plane
  therefore in the space the scattered wave $\Psi_{fin}$ 
  propogates to the solid angle $d\Omega=2\pi\sin\theta\,d\theta$.
  Returning to the  definition of probability of elastic interaction
  it is enough to multiply the wave $\Psi_{fin}$ by an amplitude:
  \begin{equation*}\label{theory.normalization wave function}
    \Phi \;=\; A(\theta)\Psi_{fin}\;,\quad \frac{d\sigma(\theta)}{d\Omega}\;=\;|\Phi|^2 \;=\; |A(\theta)|^2\;.
  \end{equation*}

  Using the well known solution of scattering problem
  \cite{Lapidus-ppt-method-Eng,Drukarev-Obedkov-Eng}
  take it as spin matrix $M(\theta)$ 
  in the Goldberger-Watson representation
  \cite{Goldberger,Goldberger-Watson-Eng}:
  \begin{equation}\label{theory.Goldberger-Watson matrix}
    M(\theta)\;=\;a\,I_{1,2}+
    b\,\sigma_{1n}\sigma_{2n}+
    c\,(\sigma_{1n}+\sigma_{2n})+
    e\,\sigma_{1m}\sigma_{2m}+
    f\,\sigma_{1l}\sigma_{2l}\;.
  \end{equation}
  $I_{1,2}$ --- multiplying of two unit 
  $2\times2$ matrices of first and second particles, 
  amplitudes $(a,\,b,\,c,\,e,f)$ 
  are the complex functions of scattering angle $\theta$ 
  and kinetic energy of incident particle in the lab system.
  New spin states are defined by the Pauli-operators
  which acts along the vectors $(\vec m,\,\vec l,\,\vec n)$: 
  \begin{equation}\label{theory.operator Pauli}
   \sigma_m = (\widehat\sigma\cdot\vec{m})\;,\quad
   \sigma_l = (\widehat\sigma\cdot\vec{l})\;,\quad   
   \sigma_n = (\widehat\sigma\cdot\vec{n})\;,
  \end{equation}
  \begin{equation*}\label{theory.vectors basic}
   \vec m=\dfrac{\vec p-\vec p\,'}{|\vec p-\vec p\,'|}\;,\quad
   \vec l=\dfrac{\vec p+\vec p\,'}{|\vec p+\vec p\,'|}\;,\quad
   \vec n=\dfrac{\vec p\times\vec p\,'}{|\vec p\times \vec p\,'|}\;.
  \end{equation*}
 
  Since the operator $\widehat{P}(\theta)$ turns the momenta
  and the matrix $M(\theta)$ changes the spin projections
  their order does not matter
  and the simultaneus action of them provides the full
  determination of scattering process:
  \begin{equation*}
   \widehat{P}(\theta)\,M(\theta) \equiv M(\theta)\,\widehat{P}(\theta)\,,\quad
   \Phi = \widehat{P}(\theta)\,M(\theta)\times\Psi_{in}\,.
  \end{equation*}

\subsection{Flip and Non-Flip parts of differential cross-section}
\begin{figure}[!ht]
   \quad\centering
   \scalebox{.25}{\includegraphics{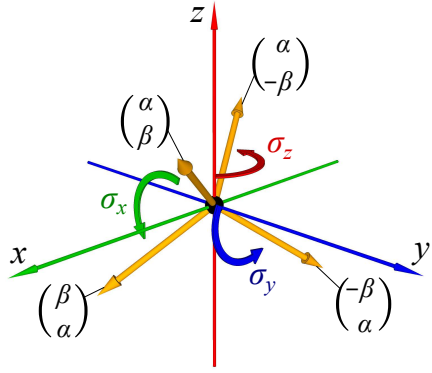}}\quad
   \caption{\small{Four directions of the fermions polarization are shown
    which are related among themselves with rotations by $180^{\,\circ}$ 
    around the axes $(x,\,y,\,z)$.
   }}\label{appendix.four vectros of polarization and their spinors}
  \end{figure}

  The arbitrarity condition of quantization axis $z$ 
  allows to direct it along the vector  $\vec{n}$,
  two other axes $x$ and $y$ will be along the vectors $\vec{m}$ and $\vec{l}$.
  The elastic scattering matrix \eqref{theory.Goldberger-Watson matrix}
  becomes more definite.
  The spin transformation is performed by the operator
  $\widehat{R}^+_t(\varphi)=E\cos\frac{\varphi}{2}+i(\widehat\sigma\cdot\vec{t}\,)\sin\frac{\varphi}{2}$
  where the $\vec{t}$ and $\varphi$ are the axis and angle of rotation.
  Since $\widehat{R}^+_t(\pi)=i\sigma_t$
  the sigma-matrices $\sigma_x,\,\sigma_y$ and $\sigma_z$
  are the rotation operators by $180^{\,\circ}$ around their axes 
  (Fig.~\ref{appendix.four vectros of polarization and their spinors}).
 
  The Pauli-operators \eqref{theory.operator Pauli}
  transform initial wave \eqref{theory.prime function modify 2}
  and give the 6 final wave functions
  which are orthogonal
  on the Bloch sphere\footnote{The orthogonality of spin-functions 
  $\chi$ and $(\widehat\sigma\cdot\vec{t}\,)\chi$ is defined as follows:
  \vspace {-2mm}
  \begin{equation*}
    \forall \quad\vec{t}\textrm{\;\,и\;\,}\chi=\binom{\cos\theta_m/2}{-e^{i\varphi_m}\sin\theta_m/2}\quad\Rightarrow\quad
    \frac{1}{4\pi}\int\limits^{2\pi}_{0}\int\limits^{\pi}_{0}
    \chi^+(\widehat\sigma\cdot\vec{t}\,)\chi\;d\theta_m\,d\varphi_m\;=\;0\,.
  \end{equation*}},
   i.e. when the both particles of beam and target are unpolarized:
  \begin{eqnarray}\label{theory.final wave functions}
   & \Psi_a=\dbinom\alpha\beta_{p}\dbinom\gamma\delta_{-p}\;,\quad
     \Psi_c'=\dbinom\alpha\beta_{p}\dbinom\gamma{-\delta}_{-p}\;,\quad
     \Psi_c''=\dbinom\alpha{-\beta}_{p}\dbinom\gamma\delta_{-p}\;, \quad \nonumber \\ [2mm]
   & \Psi_b=\dbinom\alpha{-\beta}_{p}\dbinom\gamma{-\delta}_{-p}\;,\quad
     \Psi_e=\dbinom\beta\alpha_{p}\dbinom\delta\gamma_{-p}\;, \quad
     \Psi_f=\dbinom\beta{-\alpha}_{p}\dbinom\delta{-\gamma}_{-p}\;. \quad
  \end{eqnarray}

  The action of matrix $M(\theta)$ can be considered
  with the position of spin changing of scattering particle.
  The transition $\binom\alpha\beta\to\binom\alpha\beta$
  is provided by two operators $I_{1,2}$ и
  $(\sigma_{1z}+\sigma_{2z})\,$.
  The differential cross-section of these processes is defined as follows:
  \begin{subequations}
  \begin{equation}\label{theory.Non-Flip cross section}
    \frac{d\sigma(\theta)}{d\Omega}^\textrm{Non-Flip}=|a|^2+|c|^2\;.
  \end{equation}

  The operator $(\sigma_{1z}+\sigma_{2z})$
  also performs the transition $\binom\alpha\beta\to\binom\alpha{-\beta}$
  and analogic changing is given by the operator $\sigma_{1z}\sigma_{2z}$.
  The physical sense of this transformation
  is the spin rotation by $180^{\circ}$ around the $z$-axis.
  The transitions $\binom\alpha\beta\to\binom\beta\alpha$
  and $\binom\alpha\beta\to\binom\beta{-\alpha}$ are performed by the operators
  $\sigma_{1x}\sigma_{2x}$ and $\sigma_{1y}\sigma_{2y}$ respectively
  that corresponds to the spin rotations by $180^{\circ}$ around the axes $x$ and $y$.
  The differential cross-section of these transitions 
  calculated by the formula:
  \begin{equation}\label{theory.Flip cross section}
   \frac{d\sigma(\theta)}{d\Omega}^\textrm{Flip}=|b|^2+|c|^2+|e|^2+|f|^2\;.
  \end{equation}
  \end{subequations}

\pagebreak  
\subsection{Two representations of elastic scattering}
\begin{figure}[!ht]
   \centering
   \scalebox{.37}{\includegraphics{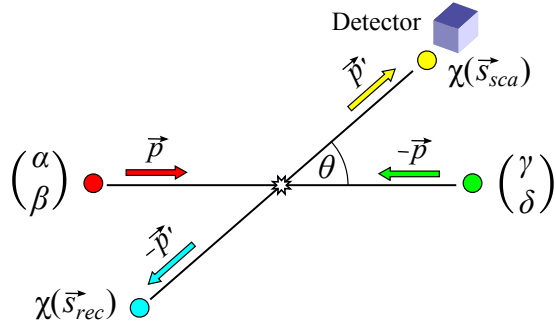}}
   \caption{\small{Elastic scattering of two identical fermions.
     One of them in the $\binom\alpha\beta$ state flies 
     from the left with momentum $\vec{p}$.
     The state of another fermion with momentum $-\vec{p}$ 
     is given by the spinor $\binom\gamma\delta$.
     Secondary particles going from the point of interaction
     with momenta $\vec{p}\,'$ and $-\vec{p}\,'$ have states
     $\chi(\vec{s}_{sca})$ and $\chi(\vec{s}_{rec})$ respectively.
   }}\label{theory.scattering and detector}
  \end{figure}

  The main task of the theory is the prediction
  what a spin states of particles will be after the interaction
  (Fig.~\ref{theory.scattering and detector}).
  If both fermions are identical
  it is impossible to predict which of them is scattered to the angle $\theta$
  and who goes in the opposite direction $\pi-\theta$.
  In this case the difinition of {\it scattered particle} is conditionally.
  \noindent

  Supposing that the initial wave function of two identical fermions 
  discribes by the form \eqref{theory.prime function modify 2}.
  The scattering matrix $M(\theta)$ determines the 6 secondary waves 
  \eqref{theory.final wave functions}
  and each of them has own amplitude.
  They form a system of wave functions 
  along which the final state vector of two particles is decomposed:
  \begin{equation}\label{theory.final full wave function}
    \Phi\;=\;\sum\limits^{6}_{i=1} A_i\Psi_{fin,\,i}\;.
  \end{equation}
  
  Anyway another wave system $\{\Psi_{fin,\,i}^\star\}$
  can be used but the transition from the first system should be unitary.
  Then the vector \eqref{theory.final full wave function}
  will receive new coordinates $\Phi=\sum A_i^\star\Psi_{fin,\,i}^\star\,$.
  If elastic ineraction is considered
  like a scattering along the momentum $-\vec{p}\,'$
  then the angle is changed $\theta\to\theta-\pi$
  and instead of matrix $M(\theta)$ 
  we need to use the matrix $M(\theta-\pi)$.
  For observer in the c.m.s. of two particles 
  it will be look as if the detector takes inverse position 
  and catches other particle
  whom named before as a recoil particle.
  But the configuration of two particles does not change
  and hence our measurement should provide the 
  same result, i.e. the vector of final state $\Phi$ \eqref{theory.final full wave function}
  which have been defined by the first approach.
  Therefore both representations of elastic scattering
  are equivalent:
  \begin{equation}\label{theory.change representation preliminary}
    \Phi\;=\;\widehat{P}(\theta)M(\theta)\times\Psi_{in}\;=\;
    \widehat{P}(\theta-\pi)M(\pi-\theta)\times\Psi_{in}\;.
  \end{equation}
  Here we take into account that $M(\theta-\pi)=M(\pi-\theta)$ 
  according to the space isotropy.
  Using the properties of operator
  \eqref{theory.properties of turn operator} it provides:
  $\widehat{P}(\theta-\pi)=\widehat{P}(\theta)\widehat{P}(\pi)$.
  But the $\widehat{P}(\pi)$ operator performs the momenta inversion
  $\widehat{P}(\pi)\times|\vec{p}_1,\vec{p}_2\rangle=|\vec{p}_2,\vec{p}_1\rangle$.
  and it is known as the Majorana operator
  $\widehat{P}_M\equiv\widehat{P}(\pi)$
  \cite{Blatt-Weisskopf}.
  Thus the expression \eqref{theory.change representation preliminary}
  allows to find:
  \begin{equation}\label{theory.change representation}
    M(\theta)\;=\;
    \widehat{P}(\pi)\times M(\pi-\theta)\;.
  \end{equation}

  Since the Bartlett operator 
  $\widehat{P}_B=\frac12(1+\widehat\sigma_1\widehat\sigma_2)$
  changes the spin states
  then the combination $\widehat{P}_M\widehat{P}_B$
  performs the full permutation:
  \begin{equation}\label{theory.Majorana operator}
    \widehat{P}_M\widehat{P}_B\times\Psi\;=\;-\Psi\quad\Rightarrow\quad
    \widehat{P}_M\;=\;-\widehat{P}_B\;=\;
    -\frac{1}{2}\left(1+\widehat\sigma_1\widehat\sigma_2\right)\;.
  \end{equation}

  Although in this analysis the fermions were considered
  but the definition of spin $\frac12\hbar$ did not use absolutely
  therefore the formula \eqref{theory.change representation}
  is sutable for any two identical particles
  with any quantum numbers.
  From this issue only the expression of Majorana operator
  should be dependent.

\section{Neutron-proton elastic scattering}\label{chapter-Theory.matrix of elastic NN-scattering}
  The $NN$-formalism repeats the method of spin physics.
  The isotopic spin $T$ is defined
  and two its projections $T_3=+\frac12$ и $T_3=-\frac12$ 
  mean the proton and neutron.
  They are entered like the isotopic spinors $\binom10$ and $\binom01$
  which can be write as $p$ and $n$ for simplification.
  Linear combinations
  \begin{equation*}\label{theory.isospin functions}
    \chi^{T}_0 = \tfrac{1}{\sqrt{2}}(p_1n_2-n_1p_2)\quad\textrm{and}\quad
    \chi^{T}_{1,\,0}  = \tfrac{1}{\sqrt{2}}(p_1n_2+n_1p_2)
  \end{equation*}
  present the nucleons in the isotopic states $T=0$ 
  and $T=1$ (with projection $T_3=0$) respectively
  and both of them are own vectors of $\widehat\tau_1\widehat\tau_2$ operator:
  \begin{eqnarray*}\label{theory.Isospin Pauli operator}
   & \widehat\tau_1\widehat\tau_2\;=\;
     \tau_{1,3}\tau_{2,3}\;+\;
     2\left[\tau_{1+}\tau_{2-}+\tau_{1-}\tau_{2+}\right]\;, \\ [2mm]
   & \tau_3 = \dbinom{1\quad0}{0\,-1}\;,\quad
     \tau_+ = \dbinom{0\quad1}{0\quad0}\;,\quad
     \tau_- = \dbinom{0\quad0}{1\quad0}\;, \\ [2mm]
   & \widehat\tau_1\widehat\tau_2\times\chi^{T}_0=-3\chi^{T}_0\;,\qquad
     \widehat\tau_1\widehat\tau_2\times\chi^{T}_{1,\,0}=+\chi^{T}_{1,\,0}\;.
  \end{eqnarray*}

  The neutron-proton wave function is build using the same scheme
  \eqref{theory.prime function modify 2}:
  \begin{equation}\label{theory.initial wave function of nucleons}
    \Psi_{in}\;=\;\dfrac{1}{\sqrt2}
    \left[n_1p_2\dbinom\alpha\beta_{\!\!1}\dbinom\gamma\delta_{\!\!2}\varphi_{12}\;-\;
    p_1n_2\dbinom\gamma\delta_{\!\!1}\dbinom\alpha\beta_{\!\!2}\varphi^*_{12}\right]\;.
  \end{equation}

  Since in the nuclear interaction
  full isospin $T$ and its projection $T_3$
  are saved\footnote{Weak processes
  \,$n\to p+e+\widetilde\nu$\, and \,$p\to n+e^++\nu$\,
  lead to a violation of isotopic invariance with probability $\sim10^{-8}$, 
  i.e. in the nuclear interactions 
  the neutron does not converted to proton at all and vice versa, 
  therefore $T=$ const  and $T_3=$ const.}
  the nucleons elastic scattering matrix has the next form:
  \begin{equation}\label{theory.nucleons scattering matrix}
    M(\theta)=M_0(\theta)
    \frac{1-\widehat\tau_1\widehat\tau_2}{4} +
    M_1(\theta)
    \frac{3+\widehat\tau_1\widehat\tau_2}{4}\;.
  \end{equation}
  Here the $M_0$ and $M_1$ are the spin matrixes 
  like \eqref{theory.Goldberger-Watson matrix}
  but present the elastic scattering
  of nucleons in pure isotopic states $T=0$ and $T=1$.
  Thus the $M_1$ describes the $np$-scattering in the state $T=1$
  and has the 5 own amplitudes $(a_1,\,b_1,\,c_1,\,e_1,\,f_1)$.
  The matrix $M_0$ with amplitudes $(a_0,\,b_0,\,c_0,\,e_0,\,f_0)$
  is intended to the $T=0$ state.
  To define the result of elastic $np$-interaction
  it is sutable to share the matrix 
  \eqref{theory.nucleons scattering matrix}
  by the direct and exchange parts:
  \begin{eqnarray}\label{theory.nucleons scattering matrix direct and exchange}
   & M(\theta)=
     \dfrac12\left(M_1(\theta)+M_0(\theta)\right) +
     \dfrac12\left(M_1(\theta)-M_0(\theta)\right)\,\widehat{P}^{\,T}_B\;, \\ 
   & \widehat{P}^{\,T}_B=\dfrac12(1+\widehat\tau_1\widehat\tau_2)\;-\;
     \textrm{isotopic exchange Bartlett operator.}\nonumber
  \end{eqnarray}

  One half of sum $\frac12(M_1(\theta)+M_0(\theta))$
  gives the simple scattering
  $np\to np$ under the angle $\theta$
  either describes the symmetric reaction $pn\to pn$.
  The second part \eqref{theory.nucleons scattering matrix direct and exchange}
  has the exchange isotopic operator $\widehat{P}^{\,T}_B$
  and adds to scattering the isospins permutation.
  Therefore the half of difference $\frac12(M_1(\theta)-M_0(\theta))$
  is the spin matrix of charge-exchange $np\to pn$ or $pn\to np$
  reactions under the angle $\theta$.

\subsection{Change representation of $np$ elastic scattering}\label{chapter-Theory.change representation of np interaction}
  Let the wave function of two nucleons is defined by the formula
  \eqref{theory.initial wave function of nucleons}.
  If the proton in the c.m.s. system of elastic $np$-interaction 
  is scattered under the angle $\theta$
  then the neutron flies in inverse direction $\pi-\theta$.
  This reaction can be presented as charge-exchange $np\to pn$ process
  when the spin states of neutron and proton are described by the matrix
  $\frac12(M_1(\theta)-M_0(\theta))$.
  Either the matrix $\frac12(M_1(\pi-\theta)+M_0(\pi-\theta))$ can be used
  to define the elastic $np\to np$ scattering under the angle $\pi-\theta$.
  Since the vector $\Phi$ of final states
  of neutron and proton can not depend 
  from our view which of these particles are registered
  the both representations are equivalent
  but can differ among themselves 
  by the set of basis wave functions of vector $\Phi$.
  According to \eqref{theory.Majorana operator}
  the transition from one representation to another
  should be performed by the Majorana operator:
  \begin{equation}\label{theory.change representation of nucleons}
    \frac12(M_1(\pi-\theta)+M_0(\pi-\theta))\;=\;
    \widehat{P}_M\times\frac12(M_1(\theta)-M_0(\theta)\,\widehat{P}^{\,T}_B\;.
  \end{equation}

  Because the nucleons have the variables of spin, isospin and space
  the full permutation is performed by three operators
  $\widehat{P}^{\phantom{x}}_M$, $\widehat{P}^{\phantom{x}}_B$
  and $\widehat{P}^{\,T}_B$:
  \begin{eqnarray*}\label{theory.Majorana operator for nucleons}
   & \widehat{P}^{\phantom{x}}_M\widehat{P}^{\phantom{x}}_B\widehat{P}^{\,T}_B\times
     \Psi\;=\;-\Psi\quad\Rightarrow\quad \\ [2mm]
   & \widehat{P}^{\phantom{x}}_M\;=\;-\widehat{P}^{\phantom{x}}_B\widehat{P}^{\,T}_B\;=\;
    -\dfrac{1}{2}\left(1+\widehat\sigma_1\widehat\sigma_2\right)\,
     \dfrac{1}{2}\left(1+\widehat\tau_1\widehat\tau_2\right)\;.
  \end{eqnarray*}

  The unitarity $|\widehat{P}^{\,T}_B|^2=1$
  transforms the \eqref{theory.change representation of nucleons} to the next:
  \begin{equation}\label{theory.change representation charge-exhcange to elastic}
    \frac12(M_1(\pi-\theta)+M_0(\pi-\theta))\;=\;
   -\widehat{P}_B\times\frac12(M_1(\theta)-M_0(\theta))\;.
  \end{equation}

  For the first this expression \eqref{theory.change representation charge-exhcange to elastic} 
  was defined in the work \cite{LL-1}.
  Before this time all amplitudes transformations were limited 
  in the frame of symmetry rules 
  \cite{Lehar-formalizm}
  using scattering angle replacing $\theta\to \pi-\theta$.
  Passing to the Goldberger-Watson
  amplitudes
  \cite{Goldberger-Watson-Eng}
  we get the canonical transformation
  \cite{Shindin-Flip-Czech,Shindin-Flip-Dubna,Shindin-Flip-PEPAN}:
  \vspace {-2mm}
  \begin{eqnarray}\label{theory.unitary-transition of nucleons}
    & c^{exch}(\theta)=\dfrac12(c_1(\theta)-c_0(\theta))=
      \dfrac12(c_1(\pi-\theta)+c_0(\pi-\theta))=c(\pi-\theta)\;, 
      \nonumber\\ [1mm]
    & \left(\begin{array}{c}
         a^{exch}(\theta) \\ [2mm]
         b^{exch}(\theta) \\ [2mm]
         e^{exch}(\theta) \\ [2mm]
         f^{exch}(\theta)
      \end{array}\right) = 
      \left(\begin{array}{cccc}
        -\frac{1}{2} & -\frac{1}{2} & -\frac{1}{2} & -\frac{1}{2} \\ [2mm]
        -\frac{1}{2} & -\frac{1}{2} & +\frac{1}{2} & +\frac{1}{2} \\ [2mm]
        -\frac{1}{2} & +\frac{1}{2} & +\frac{1}{2} & -\frac{1}{2} \\ [2mm]
        -\frac{1}{2} & +\frac{1}{2} & -\frac{1}{2} & +\frac{1}{2}
      \end{array}\right)
      \left(\begin{array}{c}
         a(\pi-\theta) \\ [2mm]
         b(\pi-\theta) \\ [2mm]
         e(\pi-\theta) \\ [2mm]
         f(\pi-\theta)
      \end{array}\right).
  \end{eqnarray}

  If the initial states of neutron and proton are defined by the spinors 
  $\binom\alpha\beta$ and $\binom\gamma\delta$ respectively
  then the amplitude $a^{exch}(\theta)$ show
  the correlation between 
  the spinor  $\binom\alpha\beta$ and final state $\chi_p$ of proton.
  In the same time the amplitude $a(\pi-\theta)$ say
  about the $\chi_p$ and spinor $\binom\gamma\delta$ correlation.
  These amplitudes can be equal only 
  if the proton spin state $\chi_p$ is symmetric
  (if at least it is average symmetric relative 
  to all directions of spin polarizations of neutron and proton)
  between the initial states $\binom\alpha\beta$ and $\binom\gamma\delta$.
  However the theory does not require this.

\pagebreak
\section{Quasielastic reaction $nd\to p(nn)$}
\subsection{Dean formula}
  The Pomeranchuk-Chew hypothesis  \cite{Chew-1,Chew-2,Pomeran-Eng,Pomeran-collection-Eng}
  as well as the A.~Migdal idea  \cite{Migdal-Eng}
  supposes to consider the charge-exchange $nd\to p(nn)$ reaction 
  using the free-nucleons scattering formalism.
  Following to this approach the initial wave function 
  of three nucleons will be presented in the next form:
  \begin{equation}\label{theory.initial wave function of three nucleons}
    \Psi_{3N}\;=\;\frac1{\sqrt{3!}}
    \begin{vmatrix}
     \;|\xi_1\rangle_1 & |\xi_2\rangle_1 & |\xi_3\rangle_1\;\; \\
     \;|\xi_1\rangle_2 & |\xi_2\rangle_2 & |\xi_3\rangle_2\;\; \\
     \;|\xi_1\rangle_3 & |\xi_2\rangle_3 & |\xi_3\rangle_3\;\;
    \end{vmatrix}\;.
  \end{equation}
  Here $\xi$ is the set of nucleon states including momentum,
  spin and isotopic spin.
  Let the $\xi_1$ expresses the beam neutron state
  and the $\xi_2$ and $\xi_3$ will be the proton and neutron
  of deuterium nucleus.
  Decomposing the determinant \eqref{theory.initial wave function of three nucleons} 
  by the elements of the first column gives:
  \begin{equation}\label{theory.initial wave function of three nucleons by one array}
    \Psi_{3N}\;=\;\dfrac1{\sqrt{3}}\sum|\xi_1\rangle_i\,
    \dfrac1{\sqrt{2}}
    \Bigl(|\xi_2\rangle_j\,|\xi_3\rangle_k-|\xi_3\rangle_j\,|\xi_2\rangle_k\Bigr)\;,
  \end{equation}
  \begin{equation*}
    \{i,j,k\} = \{1,2,3\},\;\{2,3,1\},\;\{3,1,2\}\;.
  \end{equation*}

  Since the wave functions
  \eqref{theory.initial wave function of three nucleons by one array} 
  are orthogonal
  it is enough to take one variant,
  for example $\{i$=1,\,$j$=2,\,$k$=3$\}$,
  reducing the coefficient $\frac{1}{\sqrt{3}}$.
  The incident neutron can be presented by the vector:
  $|\xi_1\rangle_1=n_1\binom\alpha\beta_{\!1}|\vec{p}_n\rangle_1$.
  The state of deutron nucleons is subject to several conditions. 
  For the first, the spins of proton and neutron should be parallel 
  because $S_d=\hbar$. Denote their states by the spinors
  $\binom\gamma\delta_2\binom\gamma\delta_3$
  here $\gamma=\cos{\lambda/2}$, $\delta=-e^{i\mu}\sin{\lambda/2}$,
  $\lambda$ and $\mu$ are zenith and asimutal angles of deutron polarization.
  For the second, the nucleons are joint by the isosinglet function
  $\chi^T_0=\frac1{\sqrt2}(p_2n_3-n_2p_3)$.
  The space part of their wave should be simmetric
  $\varphi_d(\vec{r})=\varphi_d(-\vec{r})$ 
  here $\vec{r}=\vec{r}_2-\vec{r}_3$. Therefore:
  \begin{equation}\label{theory.initial wave function of neutron and deutron}
    \Psi_{3N}\;=\;\Psi_{nd}\;=\;|\xi_1\rangle_1\,\Psi_d\;=\;
    n_1\binom\alpha\beta_{\!\!1}|\vec{p}_n\rangle_1\,
    \frac{p_2n_3-n_2p_3}{\sqrt2}\binom\gamma\delta_{\!\!2}\binom\gamma\delta_{\!\!3}\varphi_d\;.
  \end{equation}
  
  In the c.m.s. of beam neutron and proton of deuterium nucleus
  the exchange of charges and spin states is described by the matrix
  $\frac12(M_1(\theta)-M_0(\theta))\,\widehat{P}^{\,T}_B$.
  The scattering to the angle $\theta$
  is defined by the operator $\widehat{P}(\theta)$
  \eqref{theory.properties of turn operator}.
  In the deuteron system this changing is equivalent to the momentum transfer
  $\vec{p}_n\to \vec{p}_p\!'=\vec{p}_n-\vec{q}$
  that adds the exponent $e^{-\tfrac{i}{\hbar}\vec{q}\,\vec{r}_1}$ 
  to the wave function of scattered particle.
  Recoil particle obtains the
  $e^{\tfrac{i}{\hbar}\vec{q}\,\vec{r}_2}$ or $e^{\tfrac{i}{\hbar}\vec{q}\,\vec{r}_3}$
  depending from the proton number.
  \begin{equation*}\label{theory.final wave function of neutron and deutron preliminary}
      \Phi_{p(nn)}=
      \frac12(M_1(\theta)-M_0(\theta))\widehat{P}^{\,T}_B\widehat{P}(\theta)\times\Psi_{nd}=
  \end{equation*}
  \begin{equation*}
     =\sum A^{exch}_tp_1\binom{\alpha_t}{\beta_t}_{\!\!1}|\vec{p}_p\!'\rangle_1\,
     \frac{n_2n_3}{\sqrt2}\left[
        \binom{\gamma_t}{\delta_t}_{\!\!2}\binom\gamma\delta_{\!\!3}e^{\tfrac{i}{\hbar}\vec{q}\,\vec{r}_2}-
        \binom\gamma\delta_{\!\!2}\binom{\gamma_t}{\delta_t}_{\!\!3}e^{\tfrac{i}{\hbar}\vec{q}\,\vec{r}_3}
      \right]\varphi_d=
  \end{equation*}
  \begin{equation}\label{theory.final wave function of neutron and deutron}
    =\sum A^{exch}_tp_1\binom{\alpha_t}{\beta_t}_{\!\!1}|\vec{p}_p\!'\rangle_1\,
    \frac{n_2n_3}{\sqrt2}
    \left[\chi_{(-)}\cos{\tfrac{\vec{q}\,\vec{r}}{2\hbar}}\!+
    i\chi_{(+)}\sin{\tfrac{\vec{q}\,\vec{r}}{2\hbar}}\right]
    e^{\tfrac{i}{\hbar}\vec{q}\,\vec{r}_{nn}}\varphi_d\;,
  \end{equation}
  \begin{equation*}
    \vec{r}_{nn}=\frac{1}{2}(\vec{r}_2+\vec{r}_3)\;
    \textrm{--- radius-vector of two neutrons system,}
  \end{equation*}
  \begin{equation}\label{theory.summ and difference spin function of two fermions}
    \chi_{(\pm)} = \binom{\gamma_t}{\delta_t}_{\!\!2}\binom\gamma\delta_{\!\!3}\pm
    \binom\gamma\delta_{\!\!2}\binom{\gamma_t}{\delta_t}_{\!\!3}\;.
  \end{equation}
  $A^{exch}_t$ is amplitude of charge-exchange $np\to pn$ 
  elastic scattering to the angle $\theta$: \linebreak
  $A^{exch}_1=a^{exch}(\theta)=\frac12(a_1(\theta)-a_0(\theta))$ etc.
  (Tab.~\ref{theory.yields of S=0 and S=1 spin stations.table}).
  The spinors $\binom{\alpha_t}{\beta_t}$ and $\binom{\gamma_t}{\delta_t}$
  are defined by the operators which follow with their amplitudes
  \eqref{theory.final wave functions}.
  The function $\chi_{(-)}$ corresponds to the spin singlet state
  and the $\chi_{(+)}$ is a linear combination of three spin functions 
  $\chi_{1,-1}$, $\chi_{1,0}$ and $\chi_{1,+1}$.
  It is obviously that the $\chi_{(-)}$ 
  enters to the \eqref{theory.final wave function of neutron and deutron}
  with even space function $\cos{\frac{\vec{q}\,\vec{r}}{2\hbar}}$
  and the $\chi_{(+)}$ goes with odd function
  $\sin{\frac{\vec{q}\,\vec{r}}{2\hbar}}$,
  i.e. the wave of two neutrons is antisymmetric.
  Differential cross-section of $nd\to p(nn)$ reaction is defined as $|\Phi_{p(nn)}|^2$.
  Since the $\Phi_{p(nn)}$ is dependent from the $\vec{r}=\vec{r}_2-\vec{r}_3$
  it is nessesary to take the integral over the deutron space.
  If the particles were unpolarized the 
  \eqref{theory.final wave function of neutron and deutron}
  needs averaged over all spins directions:
  \begin{equation*}
     \dfrac{d\sigma(\theta)}{d\Omega}_{nd\to p(nn)}=\;\oint\overline{|\Phi_{p(nn)}|^2}\,dV\;= \\
     \vspace {-1mm}
  \end{equation*}
  \begin{equation}\label{theory.differential cross section nd->p(nn) preliminary}
     =\,\frac12\sum |A^{exch}_t|^2\left[
     \overline{|\chi_{(-)}|^2}\oint|\varphi_d|^2\cos^2{\tfrac{\vec{q}\,\vec{r}}{2\hbar}}\,dV\,+\,
     \overline{|\chi_{(+)}|^2}\oint|\varphi_d|^2\sin^2{\tfrac{\vec{q}\,\vec{r}}{2\hbar}}\,dV
     \right].
  \end{equation}

  Calculation provides the next results
  (Tab.~\ref{theory.yields of S=0 and S=1 spin stations.table}):
  \begin{equation*}
     |\chi_{(-)}|^2=2|\delta\gamma_t-\gamma\delta_t|^2\;,\quad
     |\chi_{(+)}|^2=4-|\chi_{(-)}|^2 \;,
  \end{equation*}
  \begin{equation*}
     \overline{|\delta\gamma_t-\gamma\delta_t|^2}=
     \begin{cases}
       \;0\,,\;\textrm{if}\;\,\dbinom{\gamma_t}{\delta_t}=\dbinom\gamma\delta\,, \\
       \;2/3\,,\;\textrm{if}\;\,\dbinom{\gamma_t}{\delta_t}=
       \dbinom\gamma{-\delta}\,,\;\dbinom\delta\gamma\;\textrm{or}\;\dbinom\delta{-\gamma}\,.
     \end{cases}
  \end{equation*}
  \begin{table}[!ht]
  \renewcommand{\arraystretch}{1.3}
  \parbox{0.9\textwidth}{
  \caption{\small Weight fractions of the spin states 
   $\chi_{(-)}$ and $\chi_{(+)}$
   of two neutrons system from the quasi-elastic charge-exchange reaction
  $nd\to p(nn)$.}\label{theory.yields of S=0 and S=1 spin stations.table}}
  \centering
  \begin{tabular}{|c|c|c|c|c|c|}
  \hline
  $\quad A^{exch}_t\quad$  & $ \;t\; $ &  $\quad\gamma_t\quad$  &  $\quad\delta_t\quad$
                    &  $\;\overline{|\chi_{(-)}|^2}\;$
                    &  $\;\overline{|\chi_{(+)}|^2}\;$ \\  \hline\hline
  $a^{exch}(\theta)$  & $ 1 $ &  $\gamma$    &  $\delta$    &  0    &  4    \\  \hline
  $b^{exch}(\theta)$  & $ 2 $ &  $\gamma$    &  $-\delta$   &  4/3  &  8/3  \\  \hline
  \multirow{2}*{$c^{exch}(\theta)$}  & $ 3 $ &  $\gamma$    &  $\delta$    &  0    &  4    \\  \cline{2-6}
                      & $ 4 $ &  $\gamma$    &  $-\delta$   &  4/3  &  8/3  \\  \hline
  $e^{exch}(\theta)$  & $ 5 $ &  $\delta$    &  $\gamma$    &  4/3  &  8/3  \\  \hline
  $f^{exch}(\theta)$  & $ 6 $ &   $\delta$   &  $-\gamma$    &  4/3  &  8/3  \\  \hline
  \end{tabular}
  \renewcommand{\arraystretch}{1.0}
  \end{table}

  \pagebreak
  Summing over the contributions of all amplitudes of scattering matrix 
  using the definition of Flip \eqref{theory.Flip cross section}
  and Non-Flip \eqref{theory.Non-Flip cross section}
  parts of the differential cross-section 
  of elastic $np\to pn$ charge-exchange process that gives:
  \begin{eqnarray}\label{theory.Dean formula}
    \dfrac{d\sigma(\theta)}{d\Omega}_{nd\to p(nn)}\!\!\!
    =\frac23\,\dfrac{d\sigma(\theta)}{d\Omega}^\textrm{Flip}_{np\to pn}
    \oint|\varphi_d|^2\cos^2{\tfrac{\vec{q}\,\vec{r}}{2\hbar}}\,dV\;+ \nonumber \\ [2mm]
    +\left(2\,\dfrac{d\sigma(\theta)}{d\Omega}^\textrm{Non-Flip}_{np\to pn}\!\!+
    \frac43\,\dfrac{d\sigma(\theta)}{d\Omega}^\textrm{Flip}_{np\to pn}\right)
    \oint|\varphi_d|^2\sin^2{\tfrac{\vec{q}\,\vec{r}}{2\hbar}}\,dV\;.
  \end{eqnarray}
  
  This Dean's formula\footnote{Original Dean's formula 
  use the deuteron form-factor
  $F(q)=\oint|\varphi_d|^2\cos{\tfrac{\vec{q}\,\vec{r}}{\hbar}}\,dV$:
  \begin{equation*}
    \oint|\varphi_d|^2\cos^2{\tfrac{\vec{q}\,\vec{r}}{2\hbar}}\,dV=
    \tfrac12\oint|\varphi_d|^2(1+\cos{\tfrac{\vec{q}\,\vec{r}}{\hbar}})\,dV=
    \tfrac12(1+F(q))\;,
      \vspace {-2mm}
  \end{equation*}
  \begin{equation*}
    \oint|\varphi_d|^2\sin^2{\tfrac{\vec{q}\,\vec{r}}{2\hbar}}\,dV\;=\;
    \tfrac12(1-F(q))\;.\quad
  \end{equation*}}
  for the first time was published in \cite{Dean-1,Dean-2}.
  The same result was obtained in \cite{Luboshitz-Dean,LLL}.
  The idea of method presented here 
  is that the final state of three nucleons 
  is determined by the direct action of nucleons matrix 
  without any considerations about the properties of wave function 
  of two recoil neutrons. 
  The formula \eqref{theory.Dean formula} show 
  how to use the deuterium nucleus as an  amplitudes filter
  \cite{Chew-1,Chew-2,Pomeran-Eng,Pomeran-collection-Eng}.
  If the proton is scattered to zero angle
  the transfer momentum $\vec{q}$ is going to zero too.
  It gives $\cos^2{\tfrac{\vec{q}\,\vec{r}}{2\hbar}}\approx1$,
  $\sin^2{\tfrac{\vec{q}\,\vec{r}}{2\hbar}}\approx0$.
  Therefore:
  \begin{equation}\label{theory.Dean formula simple}
    \dfrac{d\sigma(0)}{d\Omega}_{nd\to p(nn)}\!\!=\;
    \frac23\cdot\dfrac{d\sigma(0)}{d\Omega}^\textrm{Flip}_{np\to pn}\;.
  \end{equation}

  Measuring the proton yields of elastic $np\to pn$ 
  and quasi-elastic $nd\to p(nn)$ scatterings under zero angles
  we define their ratio $R_{dp}(0)$ 
  that allow to calculate the relation $r^\textrm{nfl/fl}_{np\to pn\,(0)}$
  between the Non-Flip and Flip parts 
  of differential cross-section 
  of elastic $np\to pn$ charge-exchange process:
  \begin{equation}\label{theory.non-flip to flip.formula}
    r^\textrm{nfl/fl}_{np\to pn\,(0)}=\;
    \dfrac{d\sigma(0)}{d\Omega}^{\textrm{Non-Flip}}_{np\to pn}\;\Bigl/\;
    \dfrac{d\sigma(0)}{d\Omega}^{\textrm{Flip}}_{np\to pn}
    = \;\frac{2}{3}\cdot\frac{1}{R_{dp}(0)}-1\;.
  \end{equation}

\subsection{Alternative formula}
  Let us to consider the quasi-elastic $nd\to p+nn$ charge-exchange reaction
  using another representation as the elastic $np\to np$ scattering.
  Initial wave of three nucleons is taken in the same form
  \eqref{theory.initial wave function of neutron and deutron}:
  \begin{equation}\label{theory.initial wave function of neutron and deutron full}
      \Psi_{nd}\;=\;\frac1{\sqrt3}\sum
      n_i\binom\alpha\beta_{\!\!i}|\vec{p}_n\rangle_i\,
      \frac{p_jn_k-n_jp_k}{\sqrt2}\binom\gamma\delta_{\!\!j}\binom\gamma\delta_{\!\!k}\varphi_{d,\,jk}\;,
      \vspace {-1mm}
  \end{equation}
  \begin{equation*}
     \{i,j,k\} = \{1,2,3\},\;\{2,3,1\},\;\{3,1,2\}\;,\quad
     \varphi_{d,\,jk} \;=\;
     |\vec{p}_{p^*}\rangle_j\,|\vec{p}_{n^*}\rangle_k\;.
  \end{equation*}
  
  The isotopic and spin variables are determined now by the scattering matrix
  $\frac12(M_1(\pi-\theta)+M_0(\pi-\theta))$.
  The space part of wave $|\vec{p}_n\rangle_i\,\varphi_{d,\,jk}$
  is changed by the operator $\widehat{P}(\pi-\theta)$
  which action can be shared by two steps.
  The $\widehat{P}(\pi)$ provides the momenta permutation
  $\vec{p}_{n,\,\textrm{cm}}\leftrightarrow\vec{p}_{p^*,\,\textrm{cm}}$
  and the operator $\widehat{P}(\theta)$ gives the rotation by the angle $\theta$.
  In the deutron coordinate system
  the incident neutron knocks out the proton
  and passes the momentum $\vec{p}_p\!'=\vec{p}_n-\vec{q}$.
  To the deutron wave function the exponent
  $e^{\frac{i}{\hbar}\vec{q}\,\vec{r}_i}$ is added
  by the momentum transfer $\vec{q}$
  which remains after the impact.
  If the proton has number $j$
  the action of $\widehat{P}(\pi-\theta)$ operator
  can be presented as follows\footnote{
  The Fermi-momentum $\vec{p}\,$
  defines the instantaneous transfer momentum $\vec{q}\,(\vec{p}\,)$ of $nn$-pair \linebreak
  but the observing value of $\vec{q}=\vec{p}_n-\vec{p}_p\!'$ can be supposed
  as an average $\langle\vec{q}\,\rangle$:
  \begin{equation*}
         \widehat{P}_{12}(\theta)\times|\vec{p}_n\rangle_1\,\varphi_{d,\,23}\;=\;
         |\vec{p}_n\rangle_1\oint e^{\tfrac{i}{\hbar}\vec{q}\,(\vec{p}\,)\,(\vec{r}_2-\vec{r}_1)}\,
         \Psi_H(p)\,e^{\tfrac{i}{\hbar}\vec{p}\,(\vec{r}_2-\vec{r}_3)}\,d^3\vec{p}\;=\;
         |\vec{p}_p\!'\rangle_1\;e^{\tfrac{i}{\hbar}\langle\vec{q}\,\rangle\vec{r}_2}\varphi_{d,\,23}\,,
  \end{equation*}
  where the $\Psi_H(p)$ is the momentum representation of Hulthen's $S$-wave function \cite{Hulthen}.}:
  \begin{eqnarray}\label{theory.neutron and deutron's proton permutation}
    & \widehat{P}(\theta)\widehat{P}(\pi)\times |\vec{p}_n\rangle_i\,|\vec{p}_{p^*}\rangle_j\,|\vec{p}_{n^*}\rangle_k\;=\;
      \widehat{P}(\theta)\times |\vec{p}_{p^*}\rangle_i\,|\vec{p}_n\rangle_j\,|\vec{p}_{n^*}\rangle_k\;=\nonumber \\ [2mm]
    & =\;e^{-\tfrac{i}{\hbar}\vec{q}(\vec{r}_j-\vec{r}_i)}\,
      |\vec{p}_n\rangle_j\,|\vec{p}_{p^*}\rangle_i\,|\vec{p}_{n^*}\rangle_k\;=\;
      |\vec{p}_p\!'\rangle_j\;
      e^{\tfrac{i}{\hbar}\vec{q}\vec{r}_i}\varphi_{d,\,ik}\;.
  \end{eqnarray}
  
  In the $np\to np\,$ representation 
  the neutorn is the scattered particle
  (forming with the neutron-spectator an $nn$-pair)
  and the proton is the recoil particle
  therefore the final vector will be defined by the next formula:
  \begin{eqnarray}\label{theory.final wave function of neutron and deutron in inverse representation pre-preliminary}
    \Phi_{(nn)p} &=&
        \frac12(M_1(\pi-\theta)+M_0(\pi-\theta))\widehat{P}(\pi-\theta)\times\Psi_{nd}= \nonumber \\
    &=& \frac1{\sqrt3}\sum\limits_{\{i,\,j,\,k\}}\sum\limits_{t}
        A_t\left[\frac{n_ip_jn_k}{\sqrt2}\binom{\alpha_t}{\beta_t}_{\!\!i}
        \binom{\gamma_t}{\delta_t}_{\!\!j}\binom\gamma\delta_{\!\!k}
        |\vec{p}_p\!'\rangle_j\;
        e^{\tfrac{i}{\hbar}\vec{q}\,\vec{r}_i}\varphi_{d,\,ik}\right.\;+\quad \nonumber \\
    &-&
        \left.\frac{n_in_jp_k}{\sqrt2}\binom{\alpha_t}{\beta_t}_{\!\!i}
        \binom\gamma\delta_{\!\!j}\binom{\gamma_t}{\delta_t}_{\!\!k}
        |\vec{p}_p\!'\rangle_k\;
        e^{\tfrac{i}{\hbar}\vec{q}\,\vec{r}_i}\varphi_{d,\,ij}\right]\;.
  \end{eqnarray}
  Here $A_t$ is one of fifth amplitudes of the $np\to np$ scattering to the angle $\pi-\theta$:
  $A_1=a(\pi-\theta)=\frac12(a_1(\pi-\theta)+a_0(\pi-\theta))$ etc.
  The spin states $\binom{\alpha_t}{\beta_t}$ and $\binom{\gamma_t}{\delta_t}$
  are defined by the Pauli-operators which follow with their amplitudes.
  The sum over the indexes $\{i,\,j,\,k\}$ can be ordered\footnote{To regroup 
  the elements of
  \eqref{theory.final wave function of neutron and deutron in inverse representation pre-preliminary}
  need use the cyclic substitutions: $i\to j$, $j\to k$, $k\to i$.}
  by the proton's numbers:
  \vspace {-2mm}
  \begin{eqnarray}\label{theory.final wave function of neutron and deutron in inverse representation preliminary}
     \Phi_{(nn)p} &=&
        \frac1{\sqrt3}\sum\limits_{\{i,\,j,\,k\}}\sum\limits_{t}
        A_tp_i\binom{\gamma_t}{\delta_t}_{\!\!i}|\vec{p}_p\!'\rangle_i\;\times \nonumber \\
     &\;\times&\!\!
        \frac{n_jn_k}{\sqrt2}\left[\binom\gamma\delta_{\!\!j}\binom{\alpha_t}{\beta_t}_{\!\!k}
        e^{\tfrac{i}{\hbar}\vec{q}\,\vec{r}_k} -
        \binom{\alpha_t}{\beta_t}_{\!\!j}\binom\gamma\delta_{\!\!k}
        e^{\tfrac{i}{\hbar}\vec{q}\,\vec{r}_j}\right]\varphi_{d,\,jk}\;.
  \end{eqnarray}
  
  Since all waves are orthogonal
  it is enough to take one variant $\{i$=1,\,$j$=2,\,$k$=3$\}$
  reducing the coefficient $\frac{1}{\sqrt3}$.
  Denoting the variables
  $\vec{r}=\vec{r}_2-\vec{r}_3$ and $\vec{r}_{nn}=\frac{1}{2}(\vec{r}_2+\vec{r}_3)$
  that gives:
  \vspace {-2mm}
  \begin{equation}\label{theory.final wave function of neutron and deutron in inverse representation}
     \Phi_{(nn)p}=
     \sum\!A_tp_1\binom{\gamma_t}{\delta_t}_{\!\!1}|\vec{p}_p\!'\rangle_1\,
     \frac{n_2n_3}{\sqrt2}
     \left[\chi_{(-)}\cos{\tfrac{\vec{q}\,\vec{r}}{2\hbar}}\!-
     i\chi_{(+)}\sin{\tfrac{\vec{q}\,\vec{r}}{2\hbar}}\right]
     \!e^{\tfrac{i}{\hbar}\vec{q}\,\vec{r}_{nn}}\varphi_{d,\,23}\;,
  \end{equation}
  \begin{equation}\label{theory.summ and difference spin function of two fermions 2}
     \chi_{(\pm)} = \!\!\binom\gamma\delta_{\!\!2}\binom{\alpha_t}{\beta_t}_{\!\!3}\pm
     \binom{\alpha_t}{\beta_t}_{\!\!2}\binom\gamma\delta_{\!\!3}\;.\quad
  \end{equation}

  The functions $\Phi_{(nn)p}$
  \eqref{theory.final wave function of neutron and deutron in inverse representation}
  and $\Phi_{p(nn)}$ \eqref{theory.final wave function of neutron and deutron}
  have the same structure.
  Their difference lies only in how the spin functions are defined
  \eqref{theory.summ and difference spin function of two fermions 2}
  and \eqref{theory.summ and difference spin function of two fermions}.
  The differential cross-section of $nd\to (nn)p$\, reaction
  equals to $|\Phi_{(nn)p}|^2$
  with the integration over the deutron space
  and with the averaging of all spin polarizations of three particles:
  \vspace {-2mm}
  \begin{equation*}
    \frac{d\sigma(\pi-\theta)}{d\Omega}_{nd\to (nn)p}=\;\oint\overline{|\Phi_{(nn)p}|^2}\,dV\;=
    \vspace {-1mm}
  \end{equation*}
  \begin{equation}\label{theory.differential cross section nd->(nn)p preliminary}
     =\;\frac12\sum |A_t|^2\left[
     \overline{|\chi_{(-)}|^2}\oint|\varphi_d|^2\cos^2{\tfrac{\vec{q}\,\vec{r}}{2\hbar}}\,dV\;+\;
     \overline{|\chi_{(+)}|^2}\oint|\varphi_d|^2\sin^2{\tfrac{\vec{q}\,\vec{r}}{2\hbar}}\,dV
     \right]\;.
  \end{equation}

  The spins of incident neutron and neutron-spectator
  are arbitrary directed therefore
  between the spinors $\binom{\alpha_t}{\beta_t}$ and $\binom\gamma\delta$
  the correlation is absent.
  The substitutions 
  $\binom{\alpha_t}{\beta_t}=\binom\alpha\beta$,
  $\binom\alpha{-\beta}$, $\binom\beta\alpha$ or $\binom\beta{-\alpha}$
  to the \eqref{theory.summ and difference spin function of two fermions 2}
  gives exactly values:
  \begin{equation}\label{theory.yields of S=0 and S=1 spin stations for simple np-np}
    \overline{|\chi_{(-)}|^2}=2\overline{|\gamma\beta_t-\delta\alpha_t|^2}=1
    \;,\quad
    \overline{|\chi_{(+)}|^2}=3\;.
    \vspace {-2mm}
  \end{equation}
  Then:
  \begin{eqnarray}\label{theory.differential cross section nd->(nn)p}
    \frac{d\sigma(\pi-\theta)}{d\Omega}_{nd\to (nn)p} &=&
    \frac{d\sigma(\pi-\theta)}{d\Omega}_{np\to np}\,
    \left(\frac12+\oint|\varphi_d|^2\sin^2{\tfrac{\vec{q}\,\vec{r}}{2\hbar}}\,dV\right)\;=\quad
    \nonumber \\
    &=&
    \left(1-\frac12\,F(q)\right)\frac{d\sigma(\pi-\theta)}{d\Omega}_{np\to np}\;.
  \end{eqnarray}

  The representation $np\to np\,(\pi-\theta)$ does not share
  the differential cross-section of quasi-elastic $nd\to (nn)p$ reaction
  by the Flip and Non-Flip parts of elastic $np\to np$ scattering.
  Since the transition from one representation to another is unitary
  it allows to change simultaneusly 
  in the fromula \eqref{theory.differential cross section nd->(nn)p}
  the values of angles $(\pi-\theta)\to \theta$ 
  and denoting the differential cross-sections
  by the charge-exchange reactions $nd\to p(nn)$ and $np\to pn$:
  \begin{equation}\label{theory.differential cross section nd->(nn)p change symbolics}
    \frac{d\sigma(\theta)}{d\Omega}_{nd\to p(nn)}\;=\;
    \left(1-\frac12\,F(q)\right)\frac{d\sigma(\theta)}{d\Omega}_{np\to pn}\;.
  \end{equation}

  The most interest related with the case 
  when the proton is scattered forward 
  therefore the momentum transfer $\vec{q}=\vec{p}_n-\vec{p}_p\!'$ closes to zero 
  and the deuteron form-factor $F(q)$ reaches unit. Hence:
  \begin{equation}\label{theory.differential cross section nd->(nn)p at the zero and change symbolics}
    \frac{d\sigma(0)}{d\Omega}_{nd\to p(nn)}\;=\;
    \frac12\cdot\frac{d\sigma(0)}{d\Omega}_{np\to pn}\;.
  \end{equation}
  
  Because the formulas
  \eqref{theory.differential cross section nd->(nn)p change symbolics}
  and \eqref{theory.Dean formula}
  give the same difinition of differential cross-sections
   $d\sigma(\theta)/d\Omega_{\;nd\to p(nn)}$
  removing from them the similar terms and reducing the same factors 
  it allows to define:
  \begin{equation}\label{theory.Shindin relation}
    \dfrac{d\sigma(\theta)}{d\Omega}^\textrm{Flip}_{np\to pn}\;=\;
    3\cdot\dfrac{d\sigma(\theta)}{d\Omega}^\textrm{Non-Flip}_{np\to pn}\;.
  \end{equation}

  The expression \eqref{theory.Shindin relation}
  should be regarded more as a hypothesis, 
  as an indication obtained within the impulse approximation.
  On the other hand the elastic $np\to pn$ charge-exchange reaction 
  does not depend on how the neutron is scattered on the deuteron.  
  To verify \eqref{theory.Shindin relation} it need to be done 
  the direct reconstruction of scattering amplitudes (DRSA)
  that requires the complete data set of $np$-observables
  \cite{Lapidus-ppt-method-Eng,Lehar-formalizm}.

\subsection{Equivalent formulas of two representations \\ $nd\to p(nn)\,(0)$ и $nd\to (nn)p\,(\pi)$}
  Let us to consider in more detail the case 
  when the secondary protons fly to zero angle 
  that gives $F(q)\approx1$ and allows to transform the formulas 
  \eqref{theory.differential cross section nd->p(nn) preliminary}
  and \eqref{theory.differential cross section nd->(nn)p preliminary}
  to the next view:
  \begin{equation}\label{theory.differential cross section nd->p(nn) at the zero}
    \frac{d\sigma(0)}{d\Omega}_{nd\to p(nn)} = \frac12\sum |A^{exch}_t|^2\,
    \overline{\left|\binom\gamma\delta_{\!\!2}\binom{\gamma_t}{\delta_t}_{\!\!3}-
    \binom{\gamma_t}{\delta_t}_{\!\!2}\binom\gamma\delta_{\!\!3}\right|^2}\;,\qquad
  \end{equation}
  \begin{equation}\label{theory.differential cross section nd->(nn)p at the zero}
    \frac{d\sigma(\pi)}{d\Omega}_{nd\to (nn)p} = \frac12\sum |A_t|^2\,
    \overline{\left|\binom\gamma\delta_{\!\!2}\binom{\alpha_t}{\beta_t}_{\!\!3}-
    \binom{\alpha_t}{\beta_t}_{\!\!2}\binom\gamma\delta_{\!\!3}\right|^2}\;.\qquad   
  \end{equation}
   
  These definitions of differential cross-sections
  \eqref{theory.differential cross section nd->p(nn) at the zero}
  and \eqref{theory.differential cross section nd->(nn)p at the zero}
  have the same structure.
  Their difference lies in how the spin of the neutron is defined 
  forming the singlet state $S_{nn}=0$ 
  with the neutron-spectator in both cases.  
  If the quasi-elastic $nd$-interaction is named as the $nd\to p(nn)$ reaction
  \eqref{theory.differential cross section nd->p(nn) at the zero}
  the neutron gets out as a recoil particle 
  and its spin $\binom{\gamma_t}{\delta_t}$ is presented
  by the one of four variants 
  of changed spin state of deutron's proton.
  When the $\binom{\gamma_t}{\delta_t}=\binom\gamma\delta$
  the yield of Non-Flip amplitude $a^{exch}(0)$ disappears.
  In inverse the all Flip spinors 
  $\binom\gamma{-\delta}$, $\binom\delta\gamma$ and $\binom\delta{-\gamma}$
  enter with weight $\frac43$
  (Tab.~\ref{theory.yields of S=0 and S=1 spin stations.table}).
  After the multiplying by the factor of $\frac12$ the only $\frac23$ remains
  from the Flip-part of differential cross-section
  of charge-exchange elastic reaction $np\to pn\,(0)$.
  In the other hand if the $nd\to (nn)p$ representation
  \eqref{theory.differential cross section nd->(nn)p at the zero} is used
  the beam neutron in the $np$-system 
  is considered as a particle scattered under the angle $180^{\,\circ}$.
  All spin states $\binom{\alpha_t}{\beta_t}$ 
  enter with equal weight 1 \eqref{theory.yields of S=0 and S=1 spin stations for simple np-np}
  therefore it does not give the separation of Flip and Non-Flip parts
  and from the differential cross-section of elastic reaction $np\to np\,(\pi)$ 
  only one half is remained 
  \eqref{theory.differential cross section nd->(nn)p at the zero}.

\section{Experimental data analysis of ratio $R_{dp}$}
\begin{figure}[!ht]
 \centering
  \scalebox{.37}{\includegraphics{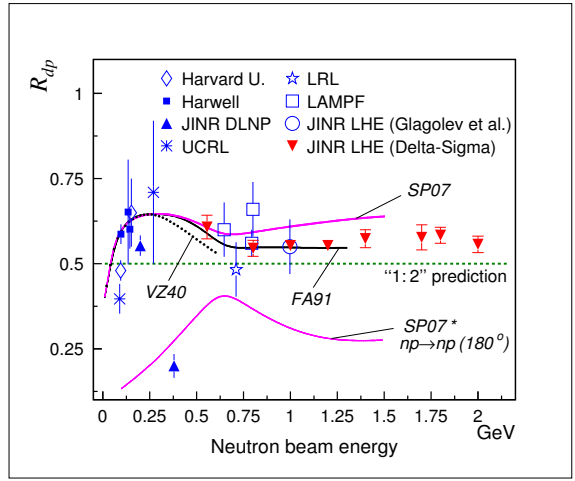}}
  \caption{\small Energydependency of the
  $R_{dp}$-ratio of protons yields 
  between quasi-elastic $nd\to p(nn)$ and elastic $np\to pn$
  charge-exchange scattering under zero.
  The PSA solutions 
  VZ40 \cite{VZ40}, FA91 \cite{FA91} and SP07 \cite{SP07} 
  are taken from SAID data base 
  as an amplitudes of $np\to np\,(\theta=\pi)$ reaction
  and transformed to the $np\to pn\,(\theta=0)$ representation 
  using \eqref{theory.unitary-transition of nucleons}.
  Values of $R_{dp}$-ratio are calculated by the fromula 
  \eqref{theory.Dean formula simple}.
  The curve SP07\,$^*$ is obtaned 
  using the Non-Flip and Flip parts 
  (\ref{theory.Non-Flip cross section}, \ref{theory.Flip cross section})
  of reaction $np\to np\,(\theta=\pi)$, i.e.
  ignoring a difference between two representations.
  Line ``1:2'' show the true approach 
  using formalism $np\to np\,(\theta=\pi)$ reaction 
  \eqref{theory.differential cross section nd->(nn)p at the zero and change symbolics}.
  }\label{Rdp.Rdp data}
 \end{figure}

  In four runs of 2003-07 
  using the Delta-Sigma Setup at LHE JINR facility 
  its were performed the measurements of $R_{dp}$-ratio between the protons yields 
  of quasi-elastic $nd\to p(nn)$ and elastic $np\to pn$ charge-exchange reactions 
  when the protons are scattered under zero angle at energies T$_n=0.5\div2.0$\,GeV  
  \cite{Shindin-Flip-Czech,Shindin-Flip-Dubna,Shindin-Flip-PEPAN,Strunov-2005,Sharov-EPJ-2009,Sharov-PoAN-2009}.
  The results of this experiment and other world data 
  are given in Att. \ref{Appendix.world data of Rdp values}
  (Tab.~\ref{Rdp.Rdp values and their errors.table},
  \ref{appendix.world Rdp values.table})
  and shown on the Fig. \ref {Rdp.Rdp data}.
  The ratio $R_{dp}$ over the energy range T$_n=0.55\div2.0$\,GeV 
  behaves like a constant on the level 0.56.
 \noindent
 
  Returning to the question about the representation choosing 
  it should be noted that the first calculations 
  of $R_{dp}$-ratio which made using the solutions of phase shift analysis (PSA) 
  were in sharp contradiction with the experiment \cite{Lehar-dubna-Eng}.
  The formula \eqref{theory.Dean formula} was used also 
  but according to \cite{BINZ-PhD} the amplitudes 
  of elastic $np\to np$ backward scattering 
  were substituted (Fig.~\ref{Rdp.Rdp data}, curve SP07\,$^*$).
  Then following to the works \cite{LL-1, Luboshitz-Dean} 
  the ratio $R_{dp}$ was recalculated with the amplitudes 
  of elastic $np\to pn$ charge-exchange scattering 
  and the same solution SP07 \cite{SP07} provided good agreement with our measurements
  \cite{Shindin-Flip-Czech,Shindin-Flip-Dubna,Shindin-Flip-PEPAN}.
  The best agreement with the experimental data 
  corresponds to the solution FA91 \cite{FA91} 
  although we should take into account 
  the approximate nature of phenomenological approaches 
  since the known $np$-data accepted for PSA are finish at the energy 1.1\,GeV.

  Our calculations by the formula 
  \eqref{theory.differential cross section nd->(nn)p at the zero and change symbolics}  
  leads to prediction $R_{dp}=1/2$ (Рис.~\ref{Rdp.Rdp data}).
  The experimental data are higher by 12\,\%
  which can be explained by several reasons.
  First, in the impulse approximation 
  the neutron-spectator is ignored
  but this is not true for large Fermi-momenta inside deuteron.
  For example, if the Hulthen's solution \cite{Hulthen} is taken 
  and $\langle P_F\rangle\approx110$\,МэВ/$c$ 
  then at energy T$_n=800$\,MeV the ratio of interaction time 
  of incident neutron and deuteron 
  to the period of its nucleons motion will about $\sim1/10$.
  It means that in 10 cases out of 100 in the reaction $nd\to p(nn)$ 
  the deuteron should be considered as a whole. \linebreak
  Secondly, starting with the meson production threshold $\sim290$\,MeV 
  the reactions of intermediate $\Delta$-resonance excitation are possible
  and these channels can change the form of 
  three nucleons wave function in the final state.
  Thirdly, in our calculations 
  the share of $D$-wave state inside deuteron ($\sim4$\,\%) was neglected.
  But if the mixing of $S$ and $D$-waves carry out 
  at once after the charge-exchange $d\to nn$ 
  it should give an additional yield too. 

  The formula 
  \eqref{theory.differential cross section nd->(nn)p at the zero and change symbolics}
  does not depend from PSA solutions 
  and it is an advantage in comparison with the formula \eqref{theory.Dean formula}
  allowing to find the discrepancy between theory and experiment.
  Because the method of obtaining of formulas \eqref{theory.Dean formula}
  and \eqref{theory.differential cross section nd->(nn)p change symbolics} is the same 
  the additive of 12\,\% should be true for both of them.
  Therefore the calculated $R_{dp}$-values
  need taken with the factor of 1.12 anyway.

\section{Conclusion}
\begin{enumerate}
 \item
  The problem of difference between two representations 
  of elastic interaction of two identical particles is investigated.
  It leads to two equivalent but not identical expressions 
  of wave function in the final state.
  The transition between these representations 
  is given by the Majorana operator.

 \item
  The method is proposed to calculation the Dean formula 
  for quasi-elastic charge-exchange reaction $nd\to p(nn)$ 
  where the wave function of proton and two neutrons in the final state 
  is defined as a direct action of nucleon scattering matrix.
  Using alternative $np\to np$ backward scattering representation 
  the new formula is established.
  The differential cross-sections of
  quasi-elastic $nd\to p(nn)$ and elastic $np\to pn$
  scattering under zero angles 
  should be related among themselves like 1:2.

 \item
  In the energy range T$_n=0.55\div2.0$\,GeV 
  the experimental ratio $R_{dp}$ 
  is like a constant on the level 0.56.
  It is more higher then calculated value $R_{dp}=1/2$ by 12\,\% 
  and related mainly with an accuracy of impulse approximation.  
\end{enumerate}
  
\thanks{A significant contribution to this study 
  was made by those who are no longer with us,
  professors L.\,N.~Strunov, F.~Lehar and V.\,L.~Lyuboshits.
  Our experiment was supported by the Russian Fund of Fundamental
  Researchers, projects No. 02-02-17129 and No. 07-02-01025.
}
  
\appendix
\include{appendix-world_rdp_data}


\bibliographystyle{my-ieeetr}
\bibliography{./bib/shindin_dubna_bib/shindin}


\end{document}

%% file: appendix-world_rdp_data.tex
 \renewcommand{\arraystretch}{1.3}
 \section{World Data}\label{Appendix.world data of Rdp values}
  \begin{table}[!ht]
  \caption{\\ Ratios $R_{dp}$ and $r^{nfl/fl}_{np\to pn}$ and their errors. \\
  Experiment Delta Sigma, 2003-07
  \cite{Shindin-Flip-Czech,Shindin-Flip-Dubna,Shindin-Flip-PEPAN,Strunov-2005,Sharov-EPJ-2009,Sharov-PoAN-2009}.}
  \label{Rdp.Rdp values and their errors.table}
  \centering
  \begin{tabular}{|c|c|c|c|c|c|c|c|c|}
  \hline
    T$_n$, ГэВ                  & 0.55  & 0.8   & 1.0   & 1.2   & 1.4   & 1.7   & 1.8   & 2.0   \\ \hline
    $R_{dp}\,(0)$               & 0.608 & 0.546 & 0.553 & 0.554 & 0.574 & 0.550 & 0.584 & 0.557 \\ \hline
    $\varepsilon$               & 0.035 & 0.024 & 0.012 & 0.010 & 0.027 & 0.034 & 0.024 & 0.024 \\ \hline\hline
    $r^{nfl/fl}_{np\to pn\,(0)}$	& 0.097 & 0.222 & 0.204 & 0.204 & 0.162 & 0.155 & 0.142 & 0.197 \\ \hline
    $\varepsilon$ 		& 0.062 & 0.053 & 0.026 & 0.023 & 0.054 & 0.074 & 0.046 & 0.052 \\ \hline
  \end{tabular}
  \end{table}

  \begin{table}[!ht]
  \caption{\\ World experimental data on ratios $R_{dp}$ and $r^{nfl/fl}_{np\to pn}$.}
  \label{appendix.world Rdp values.table}
  \centering
  \begin{tabular}{|c|c|c|c|c|c|}
    \hline
    T$_{kin}$,\,МэВ  &  $R_{dp}\,(0)$  &  $r^{nfl/fl}_{np\to pn\,(0)}$   &  Laboratory  &  Year, Ref  \\ \hline
     90  &   0.397 $\pm$ 0.044   &   0.679 $\pm$ 0.186   &  UCRL         &  1951, \cite{UCRL-90}	\\ \hline
     95  &   0.480 $\pm$ 0.030   &   0.389 $\pm$ 0.087   &  Harvard U.   &  1953, \cite{Harvard}	\\ \hline
     96  &   0.587 $\pm$ 0.029   &   0.136 $\pm$ 0.056   &  Harwell      &  1967, \cite{Harwell-2}	\\ \hline
    135  &   0.652 $\pm$ 0.154   &   0.022 $\pm$ 0.241   &  Harwell      &  1965, \cite{Harwell-1}	\\ \hline
    144  &   0.601 $\pm$ 0.057   &   0.109 $\pm$ 0.105   &  Harwell      &  1967, \cite{Harwell-2}	\\ \hline
    152  &   0.650 $\pm$ 0.100   &   0.026 $\pm$ 0.158   &  Harvard U.   &  1966, \cite{Harvard}	\\ \hline
    200  &   0.553 $\pm$ 0.030   &   0.205 $\pm$ 0.065   &  JINR LNP     &  1962, \cite{Dzhelepov-2}	\\ \hline
    270  & \;0.710 $\pm$ 0.021\; & \!\!$-0.061$ $\pm$ 0.278\; &  UCRL    &  1952, \cite{UCRL-270}	\\ \hline
    380  &   0.200 $\pm$ 0.035   &   2.333 $\pm$ 0.583   &  INP Dubna    &  1955, \cite{Dzhelepov-1-english}	\\ \hline
    647  &   0.600 $\pm$ 0.080   &   0.111 $\pm$ 0.148   &  LAMPF        &  1976, \cite{Bjork}	\\ \hline
    710  &   0.483 $\pm$ 0.080   &   0.380 $\pm$ 0.229   &  LRL          &  1960, \cite{Larsen}	\\ \hline
    794  &   0.560 $\pm$ 0.040   &   0.190 $\pm$ 0.085   &  LAMPF        &  1978, \cite{Bonner}	\\ \hline
    800  &   0.660 $\pm$ 0.080   &   0.010 $\pm$ 0.122   &  LAMPF        &  1978, \cite{Bjork}	\\ \hline
    997  &   0.550 $\pm$ 0.080   &   0.212 $\pm$ 0.176   &  JINR LHE     &  2002, \cite{Glagolev-Rdp}	\\ \hline
  \end{tabular}
  \end{table}

  \renewcommand{\arraystretch}{1.3}

%% file: npback2ndfor_eng.bbl
\begin{thebibliography}{10}

\bibitem{Lapidus-ppt-method-Eng}
S.~M. Bilen'kii, L.~I. Lapidus, R.~M. Ryndin, ``Polarized proton target in
  experiments with high-energy particles,'' {\em Physics-Uspekhi} \textbf{7},
  {\it No.}~5, pp.~721--754, 1965.

\bibitem{Drukarev-Obedkov-Eng}
G.~F. Drukarev and V.~D. Obedkov, ``Polarization phenomena in electron and
  atomic collisions,'' {\em Physics-Uspekhi} \textbf{127}, pp.~621--650, April
  1979.

\bibitem{Goldberger}
M.~L. Goldberger, Y.~Nambu, R.~Oehme, ``Dispersion relations for
  nucleon-nucleon scattering,'' {\em Ann. Phys.} \textbf{2}, {\it No.}~3,
  pp.~226--282, 1957.

\bibitem{Goldberger-Watson-Eng}
M.~L. Goldberger and K.~M. Watson, {\em Collision Theory}.
\newblock New York: Jhon Wiley and Sons, Inc., 1964.

\bibitem{Blatt-Weisskopf}
J.~Blatt and V.~Weisskopf, {\em Theoretocal Nuclear Physics}.
\newblock New York: John Wiley and Sons, Inc., 1952.

\bibitem{LL-1}
V.~L. Lyuboshitz and V.~V. Lyuboshitz, ``The nucleon charge transfer reaction
  $n+p\to p+n$ at the zero angle and the role of spin effects,'' in {\em XVII
  International Workshop on Elastic and Diffractive Scattering. Towards High
  Enegy Frontiers} (M.~Haguenauer {\em et~al.}, eds.), (Blois, Loire Valley,
  France, May 15--20, 2005), pp.~223--227, Gioi Publishers, Vietnam, 2006.

\bibitem{Lehar-formalizm}
J.~Bystrycky, F.~Lehar, P.~Wintrenitz, ``Formalizm of nucleon-nucleon
  elastic scattering experiments,'' {\em Le Journal de Physique} \textbf{39},
  {\it No.}~1, pp.~1--32, 1978.

\bibitem{Shindin-Flip-Czech}
R.~A. Shindin, D.~K. Guriev, A.~A. Morozov, {\em et~al.}, ``Separation of flip
  and non-flip parts of $np\to pn$ charge exchange at energies T$_n=0.5-2.0$
  GeV,'' {\em Eur. Phys. J. ST} \textbf{162}, {\it No.}~1, pp.~117--123, 2008.
\newblock Selected and refereed papers from conference SPIN-PRAHA 2007.

\bibitem{Shindin-Flip-Dubna}
R.~A. Shindin, D.~K. Guriev, A.~A. Morozov, {\em et~al.}, ``Separation ``flip''
  and ``non-flip'' parts of $np\to pn$ charge-exchange at energies
  T$_n=0.5-2.0$ GeV and comparison with the psa solutions,'' in {\em
  Proceedings XII Advanced Research Workshop on High Energy Spin Physics
  (DSPIN-07)} (A.~V. Efremof and S.~V. Goloskokov, eds.), (Dubna, Russia),
  pp.~353--357, Joint Institute for Nuclear Researches, Dubna, September 3-7,
  2007-08.

\bibitem{Shindin-Flip-PEPAN}
R.~A. Shindin, D.~K. Guriev, A.~A. Morozov, {\em et~al.}, ``Separation of
  differential $np\to pn$ charge-exchange cross section into flip and nonflip
  parts at T$_n=0.5-2.0$ GeV,'' {\em PEPAN Letters} \textbf{8}, {\it No.}~2 (165),
  pp.~90--96, 2011.

\bibitem{Chew-1}
G.~F. Chew, ``The inelastic scattering of high energy neutrons by deuterons
  according to the impulse approximation,'' {\em Phys. Rev.} \textbf{80},
  pp.~196--202, 1950.

\bibitem{Chew-2}
G.~F. Chew, ``A theoretical calculation of the inelastic scattering of 90-MeV
  neutrons by deuterons,'' {\em Phys. Rev.} \textbf{84}, pp.~710--716, 1951.

\bibitem{Pomeran-Eng}
I.~Ya. Pomeranchuk, ``Exchange collisions of fast neutrons with deuterons,''
  {\em Russian Academy of Sciences Reports} \textbf{LXXVII}, p.~249, 1951.

\bibitem{Pomeran-collection-Eng}
I.~Ya. Pomeranchuk, {\em {\it Collection of Scientific Works}. Elementary
  Particle Physics. Strong interactions.}, vol.~III.
\newblock ``NAUKA'', Moscow 1972.

\bibitem{Migdal-Eng}
A.~B. Migdal, ``The theory of nuclear reactions with production of slow
  particles,'' {\em Sov. Phys. JETP} \textbf{1}, pp.~2--6, 1955.
\newblock Read before the theoretical seminar at the Institute for Physical
  Problems in October of 1950.

\bibitem{Dean-1}
N.~W. Dean, ``Symmetrization effect in spectator momentum distribution,'' {\em
  Phys. Rev. D} \textbf{5}, {\it No.}~7, pp.~1661--1666, 1972.

\bibitem{Dean-2}
N.~W. Dean, ``Inelastic scattering from deuteron in the impulse
  approximation,'' {\em Phys. Rev. D} \textbf{5}, {\it No.}~11, pp.~2832--2835, 1972.

\bibitem{Luboshitz-Dean}
V.~V. Glagolev, V.~L. Lyuboshitz, V.~V. Lyuboshitz, N.~M. Piskunov,
  ``Charge-exchange breakup of the deuteron with the production of two protons
  and spin structure of the amlitude of the transfer reaction,'' {\em JINR
  Preprint} \textbf{E1-99-280}, 1999.

\bibitem{LLL}
R.~Lednicky, V.~L. Lyuboshitz, V.~V. Lyuboshitz, ``Spin effects and
  relative momentum spectrum of two protons in deuteron charge-exchange
  breakup,'' in {\em Proceedings of the XVI International Baldin Seminar on
  High Energy Physics Problems}, vol.~1, (Dubna, Russia, June 10--15, 2002),
  pp.~199--211, Dubna, 2003.
\newblock arXiv:nucl-th/0302036v1.

\bibitem{Strunov-2005}
L.~N. Strunov {\em et~al.}, ``Measurements of neutron-proton spin observables
  at $0^{\,\circ}$ using highest energy polarized $d,\,n$ probes,'' {\em Czech.
  J. Phys.} \textbf{56}, pp.~C434--C357, 2006.
\newblock Proceedings of International conference SPIN-PRAHA 2005.

\bibitem{Sharov-EPJ-2009}
V.~I. Sharov, A.~A. Morozov, R.~A. Shindin, {\em et~al.}, ``Measurements of the
  ratio $R_{dp}$ of the quasi-elastic $nd\to p(nn)$ to the elastic $np\to pn$
  charge-exchange process yields at zero proton emission angle over the
  $0.55-2.0$\,GeV neutron beam energy region,'' {\em Eur. Phys. J. A}
  \textbf{39}, pp.~267--280, March 2009.

\bibitem{Sharov-PoAN-2009}
V.~I. Sharov, A.~A. Morozov, R.~A. Shindin, {\em et~al.}, ``The ratio $R_{dp}$
  of the quasielastic $nd\to p(nn)$ to the elastic $np\to pn$
  charge-exchange-process yields at the proton emitting angle $\theta_{p,lab} =
  0^{\circ}$ over $0.55-2.0$\,GeV neutron beam energy region. experimental
  results,'' {\em Physics of Atomic Nuclei} \textbf{72}, pp.~1007--1020, 2009.

\bibitem{VZ40}
R.~A. Arndt, I.~I. Strakovsky, R.~L. Workman, ``Updated analysis of nn
  elastic scattering data to 1.6 GeV,'' {\em Phys. Rev.} \textbf{C}, {\it No.}~50,
  p.~2731, 1994.

\bibitem{FA91}
R.~A. Arndt, L.~D. Roper, R.~L. Workman, M.~W. McNaughton,
  ``Nucleon-nucleon partial-wave analysis to 1.6 GeV,'' {\em Phys. Rev.}
  \textbf{D}, {\it No.}~45, p.~3995, 1992.

\bibitem{SP07}
R.~A. Arndt, W.~J. Briscoe, I.~I. Strakovsky, R.~L. Workman, ``Updated
  analysis of nn elastic scattering to 3 GeV,'' {\em Phys. Rev.} \textbf{C},
  {\it No.}~76, p.~025209, 2007.

\bibitem{Lehar-dubna-Eng}
F.~Lehar, ``The end of pretty fabula and unexpected results on the Nuclotron,'' {\em
  newspaper ``Dubna''}, pp.~2--3, 2 december 2005.

\bibitem{BINZ-PhD}
R.~Binz, {\em Untersuchung der spinabh\"{a}ngigen Neutron-Proton Wechselwirkung
  im Energiebereich von 150 bis 1100\,MeV}.
\newblock PhD thesis, Fakult\"{a}t f\"{u}r Physic der Universit\"{a}t Freiburg
  i. Br., 1991.

\bibitem{Hulthen}
L.~Hulth\'{e}n and M.~Sugavara, {\em The Two-Nucleon Problem}, vol.~39,
  pp.~32--33, 76, 92.
\newblock Springer-Verlag, Berlin, 1957.

\bibitem{UCRL-90}
N.~Powell {\em Preprint UCRL 1191}, 1951.

\bibitem{Harvard}
J.~A. Hofman and K.~Strauch, ``On the interaction of 95-MeV protons with d, li,
  be, c, al, cu, pb nuclei,'' {\em Phys. Rev.} \textbf{90}, pp.~449--460,
  1953.

\bibitem{Harwell-2}
A.~Langsford {\em et~al.}, ``Fast forward neutron production in the d(p, n)2p
  reaction for 95.7 and 143.9 MeV protons,'' {\em Nucl. Phys.} \textbf{A},
  {\it No.}~99, pp.~246--268, 1967.

\bibitem{Harwell-1}
E.~J. Esten, T.~C. Griffith, G.~J. Lush, A.~J. Metheringham, ``Inelastic
  proton-deuteron scattering at 135 MeV,'' {\em Rev. Mod. Phys.} \textbf{A},
  {\it No.}~3, p.~533, 1965.

\bibitem{Dzhelepov-2}
V.~P. Dzhelepov, ``Recent investigations on nucleon-nucleon scattering at the
  dubna synchrocyclotron,'' in {\em International Conference on High-Energy
  Physics at CERN}, (Geneva), pp.~19--23, 4-11 July 1962.

\bibitem{UCRL-270}
J.~R. Cladis, J.~Hadley, W.~N. Hess, ``Fast protons from 270-MeV n-d
  collisions,'' {\em Phys. Rev.} \textbf{86}, pp.~110--117, 1952.

\bibitem{Dzhelepov-1-english}
V.~P. Dzhelepov {\em et~al.}, ``Experimental investigation of neutron-nucleon
  and neutron-deuteron interaction in the energy region 380–590 MeV,'' {\em IL
  Nuovo Cimento} \textbf{III}, {\it No.}~1, pp.~61--79, 1956.
\newblock Reported at a sessions of the Academy of Sciences of the USSR,
  December 17, 1954.

\bibitem{Bjork}
C.~W. Bjork, P.~J. Riley, B.~E. Bonner, {\em et~al.}, ``Neutron spectra at
  $0^{\,\circ}$ from p-p and p-d collisions at 647 and 800 MeV incident
  energies,'' {\em Phys. Lett.} \textbf{B}, {\it No.}~63, pp.~31--34, 1976.

\bibitem{Larsen}
R.~R. Larsen, ``Neutron-proton scattering and the determination of the
  pion-nucleon coupling constant,'' {\em IL Nuovo Cimento} \textbf{XVIII},
  {\it No.}~5, pp.~1039--1042, 1960.

\bibitem{Bonner}
B.~E. Bonner, J.~E. Simmons, J.~M. Wallace, {\em et~al.}, ``Quasielastic charge
  exchange in $n\,^2\textrm{H}\to pnn$ at 794 MeV,'' {\em Phys. Rev.}
  \textbf{C}, {\it No.}~17, pp.~664--670, 1978.

\bibitem{Glagolev-Rdp}
V.~V. Glagolev {\em et~al.}, ``Spin-dependent $np\to pn$ amplitude estimated
  from $dp\to ppn$,'' {\em Eur. Phys. J. A} \textbf{15}, {\it No.}~4, pp.~471--475,
  2002.

\end{thebibliography}
